\documentclass[fleqn,usenatbib]{mnras}

\usepackage{graphicx}
\usepackage{amsmath}
\usepackage{amssymb}
\usepackage{newtxtext,newtxmath}

\usepackage{amsmath,amstext}
\usepackage[T1]{fontenc}
\usepackage{aecompl}
\usepackage{comment}
\usepackage[figure,figure*]{hypcap}
\usepackage{multirow}
\usepackage{amsmath}
\usepackage{nccmath}
\usepackage{pdflscape}
\usepackage{rotating}
\usepackage{soul}
\usepackage{booktabs}
\setstcolor{red}
\usepackage{float}
\usepackage{longtable}
\usepackage{dcolumn}
\usepackage{orcidlink}
\usepackage{xspace}

\newcommand{\exofs}{EMOs\xspace}
\newcommand{\exof}{EMOs\xspace}
\newcommand{\supa}  {$^\mathrm{a}$}
\newcommand{\supb}  {$^\mathrm{b}$}
\newcommand{\supc}  {$^\mathrm{c}$}
\newcommand{\supd}  {$^\mathrm{d}$}

\newcommand{\dr}{DR\,21\xspace}
\newcommand{\gf}{G5.89\xspace}
\newcommand{\rahms}[4]{$#1^{\rm h}#2^{\rm m}#3\mbox{$^{\rm s}\mskip-7.6mu.\,$}#4$} %% \dechms{04}{14}{12}{9198} = RA en formato 04^h 14^m 12^s .9198
\newcommand{\decdms}[4]{$#1^{\circ}#2'#3\mbox{$''\mskip-7.6mu.\,$}#4$} %% \decdms{28}{12}{12}{199} = Dec en formato 28^o 12' 12'' .199

\title[VLA observations of DR\,21 and G5.89]{A VLA search for compact radio sources in the explosive molecular outflows DR\,21 and G5.89}

\author[V. Yanza et al.]{
Vanessa~Yanza,$^{1}$\thanks{E-mail: v.yanza@irya.unam.mx}
Sergio~A.~Dzib,$^{2}$
Aina Palau,$^{1}$
William J. Henney,$^{1}$
Luis F. Rodr\'{\i}guez,$^{1}$
Luis A. Zapata,$^{1}$
\\
$^{1}$Instituto de Radioastronom\'ia y Astrof\'isica, Universidad Nacional Aut\'onoma de M\'exico,
              Antigua Carretera a P\'atzcuaro 8701,
              Ex-Hda. San Jos\'e de la Huerta,\\
              58089 Morelia, Michoac\'an, M\'exico\\
$^{2}$ Max-Planck-Institut f\"ur Radioastronomie,  Auf dem H\"ugel 69, D-53121 Bonn, Germany
}

\date{Accepted XXX. Received YYY; in original form ZZZ}

\pubyear{\the\year{}}

\begin{document}
\label{firstpage}
\pagerange{\pageref{firstpage}--\pageref{lastpage}}
\maketitle

\begin{abstract}
We present high-angular-resolution ($\sim0\rlap{.}''1$) VLA Ku-band (12--18 GHz) observations of two explosive molecular outflows (\exofs), DR\,21 and G5.89, in a search for runaway stars related to these explosive events. In DR\,21, we identified 13 compact radio sources (CRS), 9 located in the \dr core and near the CO streamer ejection region. The radio properties of the CRSs show that three are nonthermal radio emitters, likely magnetically active stars, while the nature of the remaining CRSs cannot be conclusively identified. All detected CRSs are good candidates for follow-up proper motion studies to confirm whether they are runaway stars. We also identify multiple ionized arc-shaped structures that can be fitted with parabolas whose symmetry axes converge to a position coincident with CRSs \#11, raising the possibility that this source is the main ionizing star. A re-analysis of the 18 molecular outflow streamers refines the center of the explosive event, which aligns closely with the position indicated by the arcs convergence point, supporting a common stellar origin for the \exofs and the \ion{H}{II}-region.  In G5.89, the observations reveal a shell with a square-like morphology. The strong extended emission from this \ion{H}{II} region prevents the detection of weak compact radio sources inside the shell; only two were identified well beyond the shell, and a single parabolic arc was fitted within this region. Overall, arc structures in ionized regions seem to be good tracers of the origin of the ionizing sources.
\end{abstract}


\section{Introduction}

\begin{figure*}%[hbt!]
\includegraphics[width=0.82\linewidth %, trim= 0 150 0 0,clip
]{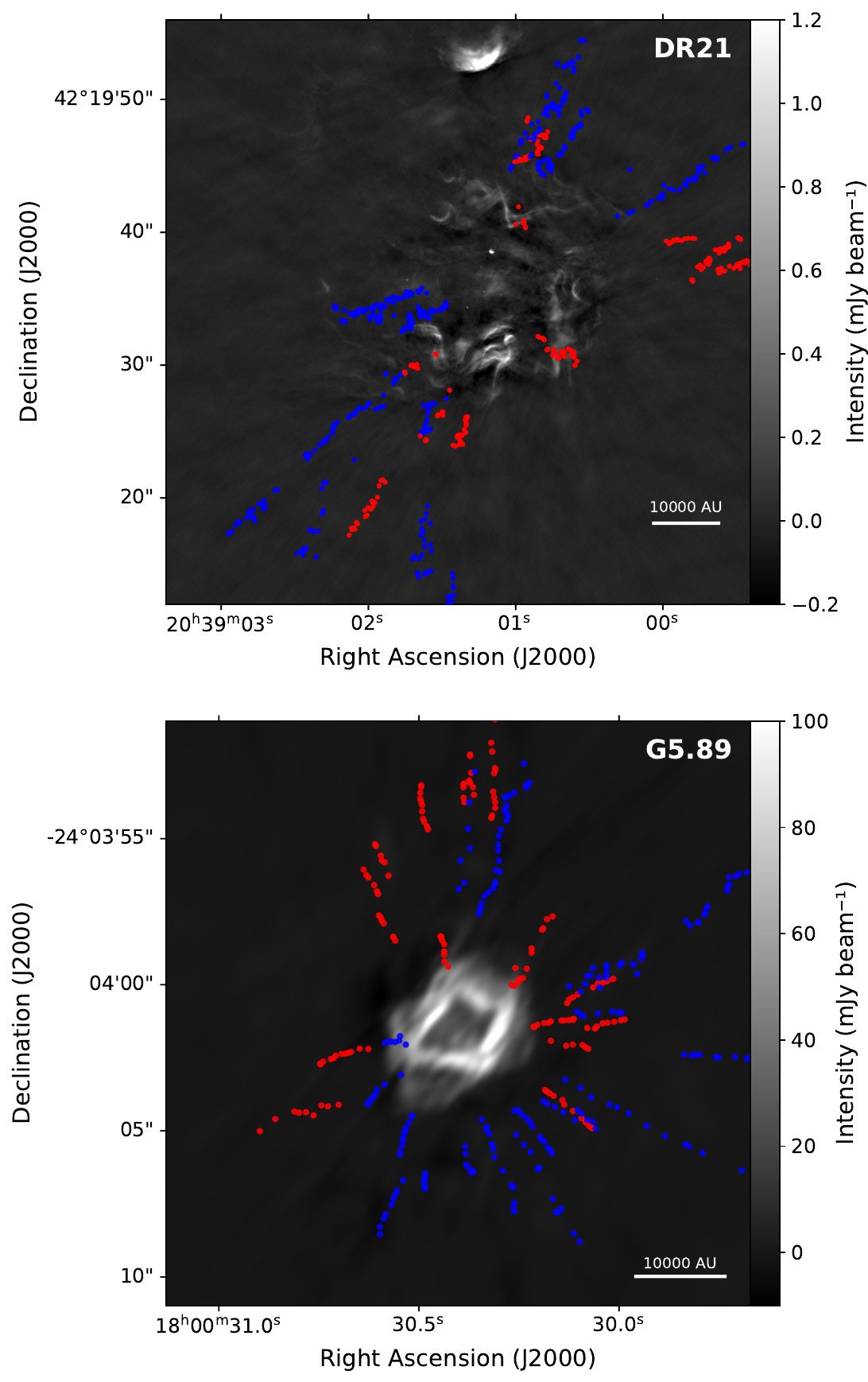}
\caption{VLA radio images of \dr (top) and \gf (bottom) at K$_u$ band (12--18\,GHz). Dots mark the blue- and redshifted streamers detected with ALMA by \citet{guzman2024} and \citet{zapata2020} respectively.}\label{fig:streams}
\end{figure*}

In recent decades, astronomers have identified a new type of outflow known as explosive molecular outflows (\exofs) in several massive star-forming regions \citep[e.g.,][]{allen1993, zapata2013, zapata2019}. Unlike common stellar outflows, which are driven by accretion processes and typically exhibit a clear bipolar structure, \exofs are characterized by their unique morphology, higher energy output, and enigmatic formation mechanism \citep[see][for a detailed discussion]{zapata2017}. Protostellar outflows generally possess kinetic energies ranging from $10^{43}$ to $10^{46}$\,ergs \citep[][]{bachiller1996, bally2016}. In contrast, \exofs eject material from a single point in multiple directions, resulting in a network of {\it finger-like} filaments distributed almost isotropically around the ejection point. These outflows are significantly more energetic, with kinetic energies ranging from $10^{47}$ to $10^{49}$\,ergs \citep{zapata2017}. While the precise formation mechanism of \exofs remains a mystery, current evidence suggests that they may result from the dynamic disruption of a compact stellar cluster. One possible scenario is the decay of a compact, non-hierarchical multiple star system into an ejected binary and two runaway stars, in which the molecular outflow is the result of the high-speed expulsion of debris from a disrupted circumstellar disk during the final stellar encounter \citep{bally2011}.

The Orion BN/KL region was the first site where \exofs were discovered \citep{allen1993} and subsequently characterized in detail \citep[e.g.,][]{zapata2009, bally2011}. Early observations at radio and millimeter wavelengths revealed that two compact radio sources (i.e., unresolved or slightly resolved sources) associated with young massive stars, known as the BN object and source I, exhibited unusual motions, moving away from each other at high velocities \citep{plambeck1995, rodriguez2005}. Subsequent studies confirmed these peculiar motions and identified additional stars in the vicinity with similar behavior \citep{gomez2005, gomez2008, dzib2017, luhman2017}. Remarkably, all these stars appear to be moving away from the same origin point, which coincides with the center of the outflows and CO streamer. The ejection time for these stars, approximately 500 years ago, coincides with the estimated ejection time for the CO streamers and outflows \citep{zapata2009, rodriguez2017, rodriguez2020}. The relative proximity of the BN/KL region in Orion, at a distance of $\sim$390 pc \citep{kounkel2017}, has facilitated numerous high-resolution studies. In contrast, other recently discovered \exofs are located at distances an order of magnitude greater. This makes it challenging to characterize stellar motions in these regions due to the distance and the inherent extinction in star-forming regions.

Radio observations offer distinct advantages in the search for stars with high velocities associated with \exofs. These observations provide high angular resolution and precise astrometric results, allowing astronomers to filter out extended emission from surrounding material. Unlike observations at other wavelengths, radio waves are unaffected by dust obscuration, a critical benefit in star-forming regions where stars are often deeply embedded. Leveraging these advantages, we present the first census of compact radio sources in the confirmed \exofs \dr \citep{zapata2013, guzman2024} and G5.89--0.39 \citep{zapata2019, zapata2020}.

\subsection{Target regions: \dr and \gf}
The \dr compact \ion{H}{II} region (also known as \dr Main, \dr hereafter) is located in the Cygnus X complex and at a distance of 1.50$^{+0.08}_{-0.07}$ kpc \citep{rygl2012}, and was first identified as a potential \exofs host by \citet{zapata2013}. While the infrared outflow initially appeared bipolar with a wide opening angle, several factors suggest it is not a typical stellar outflow but an \exofs. The outflow mass exceeds 3000\,M$_\odot$, with a kinetic energy on the order of $10^{48}$\,ergs, and it exhibits a luminosity of approximately 2000 L$_\odot$ in the 2\,$\mu$m band \citep{garden1991, garden1992}. \citet{zapata2013} identified CO filaments widely distributed around DR21, all pointing back to its core and following a Hubble law, suggesting that \dr is related to \exof. \citet{zapata2013} located the origin of the \exofs
at the position $\alpha=$\rahms{20}{39}{01}{1}$\pm5''$
and $\delta=$\decdms{+42}{19}{37}{9}$\pm5''$. However, \citet{skretas2023} found no clear evidence of \exof using \ion{HCO}{$^+$} observations from IRAM 30m and NOEMA, opening a debate about the nature of this source. The likely cause of this discrepancy is the use of \ion{HCO}{$^+$} as a tracer for EMOs. This molecule has a significantly higher excitation temperature and different abundances compared to the typical outflow tracer, CO.
More recently, \citet{guzman2024} employed sensitive ALMA observations to uncover CO filaments radially distributed around the \dr core and suggested a dynamical age of 8600 years. These authors concluded that the morphology and kinematics of these filaments confirm the explosive nature of the outflows. Finally, they located a different position for the origin of the \exofs
%at the position 
at $\alpha=$\rahms{20}{39}{00}{8}$\pm3''$ and $\delta=$\decdms{+42}{19}{37}{6}$\pm1''$, though this position is consistent within uncertainties with the position suggested by \citet{zapata2013}.% \citep{guzman2024}.

The ultracompact \ion{H}{II} region G5.89--0.39, also known as W28 A2 (hereafter \gf), is situated at a distance of $2.99^{+0.19}_{-0.17}$\,kpc \citep{sato2014}. \citet{zapata2019} first proposed \gf as an \exof, largely due to the high energy associated \citep[$\sim10^{47}$\,ergs][]{acord1997, klaasen2006} and the discovery of CO filaments pointing back to its center. Subsequent ALMA observations by \citet{zapata2020} revealed additional CO filaments widely distributed around \gf, confirming its classification as an \exof. These authors also noted that massive stars detected in the infrared at the edges of the \ion{H}{II} region may have participated in the event that triggered the \exof, which could explain their current positions outside the center of the \ion{H}{II} region. This is consistent with previous suggestions \citep[e.g.][]{feldt2003}.

%--------------------------------------------------------------------
\section{Observations}

The observations were taken with the {\it Karl G. Jansky} Very Large Array (VLA) of the National Radio Astronomy Observatory (NRAO), % VLA 23A-243
in its A configuration. The Ku-band (12--18 GHz) receiver was used covering the full-band (FB). The 
observations were in the semi-continuous mode; the entire band was split into 
48 spectral windows (SPWs) of 128\,MHz and each SPW consisted of 64 channels 
of 2\,MHz. Each target, \dr and \gf, was observed twice with a separation
of two days; thus, four observing sessions were obtained. 

Each session started with a scan on the amplitude calibrator J1331+3030 
(also known as 3C\,286), followed by intercalating scans of one minute in the gain  
calibrator and 3.5 minutes in the target source. The gain calibrators were J2007+4029 (for \dr) and J1820--2528 (for \gf). For a proper instrumental set, we follow the NRAO recommendations and add pointing and requantizer scans  before the amplitude calibrator scan and prior to the first 
scan on the gain calibrator. The reference pointing scan is done using the X-band (8--12 GHz) receivers.   The session duration was one hour, where about 23 minutes were effectively spent on the target.

Data sets were calibrated and edited using the calibration pipeline
provided by the NRAO as part of the Common Astronomy Software 
Applications \citep[CASA, ][]{casa2022}. The pipeline includes flagging 
of flawed data and obtaining gain and phase calibration\footnote{NRAO 
gives a detailed description of the pipeline data editing and calibration at \url{https://science.nrao.edu/facilities/vla/data-processing/pipeline}.}.

For \dr\ after data calibration, a series of images were done in CASA using the task {\it tclean}. First, the FB of the two sessions were combined and imaged together, which allowed the most sensitive image to extended radio sources (see Fig.~\ref{fig:streams}). Second, the two combined sessions and each session were imaged using baselines above 300\,k$\lambda$, allowing to filter out radio sources with sizes larger than 0\rlap{.}$''$69. Each \dr session was imaged in Stokes V, and they were divided into two sub-bands; the lower-side band (LSB) and the upper-side band (USB) images were obtained to cover the frequency ranges from 12 to 15 GHz and from 15 to 18\,GHz, respectively. In all cases, \dr\ images were constructed using square pixels with a size of 0\rlap{.}$''$02, image size of 2700 pixels per side, and a uniform weighting scheme. 

For \gf\ region, self-calibration was applied since its extended emission generates high rms noise in the surrounding areas, preventing the detection of compact sources. 
%rms noise of 580 and 280 $\mu$Jy beam$^{-1}$ for images with all baselines and baselines above 300\,k$\lambda$ respectively. 
After self-calibration, the FB images combining the two observing sessions were produced using all baselines and using only baselines longer than 300\,k$\lambda$. The application of self-calibration reduced the noise by a factor of approximately three compared to the images obtained without it. As in the case of \dr, LSB and USB images covering the 12--15\,GHz and 15--18\,GHz frequency ranges, respectively, were also generated using baselines longer than 300\,k$\lambda$. All \gf\ images using $>$300,k$\lambda$ were produced with uniform weighting, a pixel size of 0\rlap{.}$''$05, and an image size of 5600 pixels per side. The FB image with full ($u,v$) coverage was obtained using Briggs weighting with a robust parameter of 2, a pixel size of 0\rlap{.}$''$02, and an image size of 14400 pixels, to better trace the shell morphology of \gf\ and cover the full primary beam. In this case, individual-epoch images were not generated due to their low sensitivity; therefore, only the combined images were used. The final properties of all images obtained from the present observations are summarized in Table~\ref{tab:obs}.

%After data calibration, a series of images were done in CASA using the task {\it tclean}.  First, the FB of the two sessions were combined and imaged together, which allowed the most sensitive images to extended radio sources (see Fig.~\ref{fig:main}).
%%% Falta incluir las propuiedades de los mapas completos. 
%Second, the two combined sessions and each session were imaged for both targets using baselines above 300 k$\lambda$, allowing to filter out radio sources with sizes larger than 0\rlap{.}$''$69. After a quick visual analysis of these images, we decided to proceed with further imaging only for \dr\footnote{The reason for the decision will become evident in the next section.} by using only baselines above 300\,k$\lambda$. Each \dr session was imaged in Stokes V, and they were divided into two sub-bands; the lower-side band (LSB) and the upper-side band (USB) images were obtained to cover the frequency ranges from 12 to 15 GHz and from 15 to 18\,GHz, respectively. In all cases, the images were constructed using square pixels with a size of 0\rlap{.}$''$02, image size of 2700 pixels per side, and a uniform weighting scheme. The final properties of all produced images from the present observations are listed in Table~\ref{tab:obs}.

 \begin{table}
\small
\begin{center}
 \caption{VLA observations and image properties.}
    \begin{tabular}{ccccc}
         \hline
         \hline
Target & {Date} & Sub- & {Synthesized beam}& {rms noise}  \\
         & (07/2023) &band  &($\theta_{\rm maj.}\times\theta_{\rm min.}$; P.A.) & ($\mu$Jy\,bm$^{-1}$) \\
\hline
DR21&2 & FB &$0\rlap{.}''137\times0\rlap{.}''083$; $-86\rlap{.}^{\circ}5$  & 24  \\
& & FB* &$0\rlap{.}''137\times0\rlap{.}''083$; $-86\rlap{.}^{\circ}5$  & 13  \\
& &LSB &$0\rlap{.}''160\times0\rlap{.}''095$; $-87\rlap{.}^{\circ}1$  & 33  \\
& &USB &$0\rlap{.}''136\times0\rlap{.}''078$; $-89\rlap{.}^{\circ}3$  & 29  \\
&4 & FB &$0\rlap{.}''112\times0\rlap{.}''082$; $76\rlap{.}^{\circ}4$  & 24  \\
& & FB* &$0\rlap{.}''112\times0\rlap{.}''082$; $76\rlap{.}^{\circ}4$  & 15  \\
& & LSB &$0\rlap{.}''138\times0\rlap{.}''100$; $75\rlap{.}^{\circ}6$  & 34  \\
& &USB &$0\rlap{.}''112\times0\rlap{.}''082$; $76\rlap{.}^{\circ}4$  & 29  \\
&2+4 & FB& $0\rlap{.}''113\times0\rlap{.}''078$; $86\rlap{.}^{\circ}4$  & 19  \\
& & FB* &$0\rlap{.}''113\times0\rlap{.}''078$; $86\rlap{.}^{\circ}4$  & 11  \\
& & LSB &$0\rlap{.}''139\times0\rlap{.}''093$; $87\rlap{.}^{\circ}0$  & 23  \\
& &USB &$0\rlap{.}''113\times0\rlap{.}''075$; $85\rlap{.}^{\circ}4$  & 22  \\
& & FB$^{\rm a}$& $0\rlap{.}''116\times0\rlap{.}''080$; $86\rlap{.}^{\circ}5$  & 20  \\
\hline
G5.89&2+4& FB&$0\rlap{.}''309\times0\rlap{.}''093$; $-34\rlap{.}^{\circ}3$& 44\\
&& LSB&$0\rlap{.}''307\times0\rlap{.}''111$; $-34\rlap{.}^{\circ}5$& 91\\
&& USB&$0\rlap{.}''321\times0\rlap{.}''089$; $-32\rlap{.}^{\circ}9$& 81\\
&& FB$^{\rm a}$&$0\rlap{.}''527\times0\rlap{.}''203$; $-39\rlap{.}^{\circ}2$& 692\\
         \hline
    \end{tabular} \label{tab:obs}
\end{center}
Notes: Images were reconstructed using baselines $>300\,$k$\lambda$, except when stated otherwise. {Noise was measured near the center of the images, outside the shells.} \\
$^{\rm a}$ Image created using all the array baselines.\\
* Stokes V image.
\end{table}

\section{Results}

\begin{figure*}%[hbt!]
\includegraphics[width=1.0\linewidth
%, trim= 0 150 0 0,clip
]{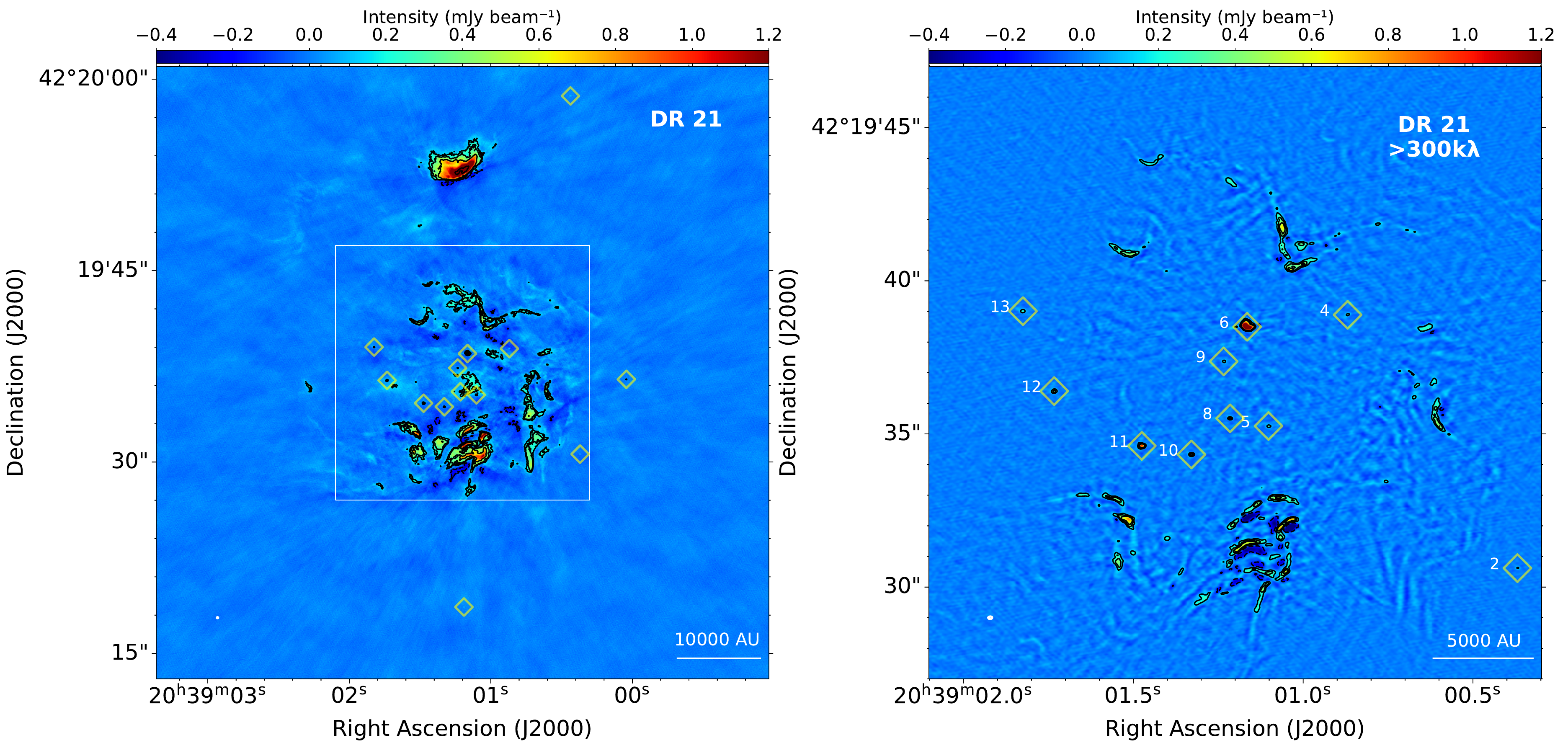}
\caption{ VLA radio images of \dr at a mean frequency of 15\,GHz. {\it Left:} Image combines the data of the observed session and use the full band and all baselines. Contour levels are at --3, 3, 5, 10, 30, and 60 times the noise level near to the arcs, 56\,$\mu$Jy beam$^{-1}$. The white square indicates the area shown in the right panel. {\it Right:} Image obtained combining the data of the observed session, using the full band and baselines $>300\,{\rm k}\lambda$. Contour levels are at --5 (red), 5, 10 and 15 times the noise level close to the arcs, 30\,$\mu$Jy beam$^{-1}$. The diamonds indicate the position of detected compact radio sources. The corresponding synthesized beam is shown in the bottom left corner of each panel.
%VLA radio images of \dr (left) and \gf (right) at a mean frequency of 15\,GHz. {\it Top panels:} These images combine the data of the observed session per target and use the full band and all baselines. Contour levels are at --3 (red), 3, 5, 10, 30, and 60 times the noise levels of the images; 56\,$\mu$Jy beam$^{-1}$ and 582\,$\mu$Jy beam$^{-1}$ for \dr and \gf, respectively. White squares indicate the area shown in the bottom panels. Additionally, the radio sources named A, B, and C are marked with ellipses. {\it Bottom panels:} These images combine the data of the observed session per target and use the full band and only baselines $>300\,{\rm k}\lambda$. Contour levels are at --3 (red), 3, 5, 10, 30, and 60 times the noise levels of the images; 19\,$\mu$Jy beam$^{-1}$ and 280\,$\mu$Jy beam$^{-1}$ for \dr and \gf, respectively. In the case of DR21 images, the diamonds indicate the position of detected compact radio sources.
}\label{fig:main}
\end{figure*}

\begin{figure*}%[hbt!]
\includegraphics[width=1.0\linewidth
%, trim= 0 150 0 0,clip
]{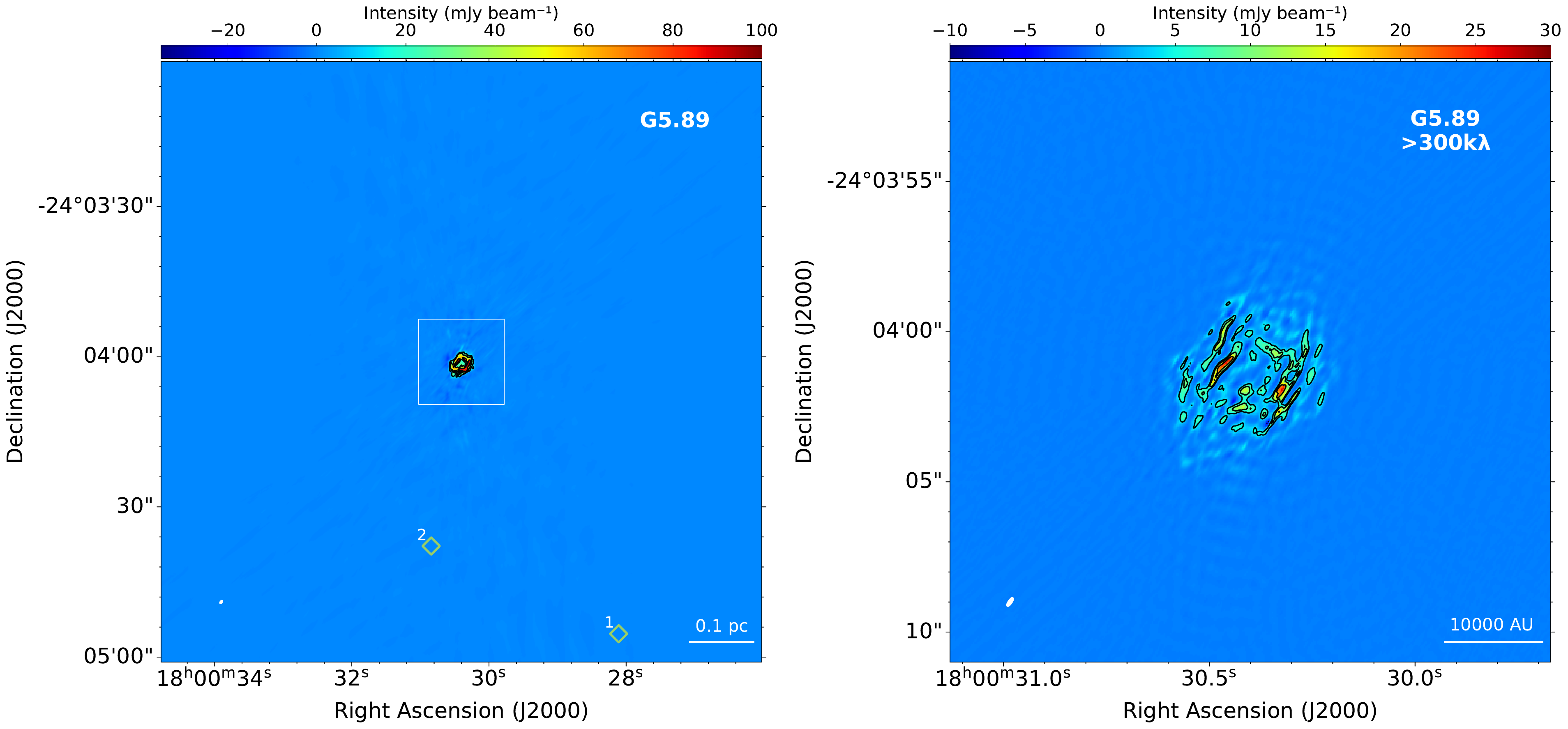}
\caption{VLA radio images of \gf\ at a mean frequency of 15\,GHz. {\it Left:} Image combines the data of the observed session and use the full band and all baselines. Contour levels are at --25, 25, 50,100 and 200 times the noise level of the image, 692\,$\mu$Jy beam$^{-1}$. The white square indicates the area shown in the right panel. The diamonds indicate the position of detected compact radio sources. {\it Right:} Image obtained combining the data of the observed session, using the full band and baselines $>300\,{\rm k}\lambda$. Contour levels are at --100, 100, 200 and 300 times the noise level, 44\,$\mu$Jy beam$^{-1}$. The corresponding synthesized beam is shown in the bottom left corner of each panel.
}\label{fig:g5}
\end{figure*}

Figure~\ref{fig:streams} shows our \dr\ and \gf\ FB images 
overlaid with the CO streamers reported by \citet{zapata2020} and \citet{guzman2024}.
This images are the FB images using all baselines and combining the two epochs.

In both 
images, extended emission dominates the resulting radio maps and the 
streamers start to be visible outside the core and shell structure observed in each region.
In the case of \dr, several arc-shaped structures are present, as 
well as some compact radio sources (CRSs), that are significantly weaker 
than the extended emission. For G5.89, the square-like shell structure of this source is  
recovered. At first inspection, no CRSs can be identified clearly.

To better identify CRSs, we produced
maps using only baselines larger than 300\,k$\lambda$ for \dr\ and \gf. The images combining the two sessions,
with and without the $uv$-range  restriction are
shown in Fig.~\ref{fig:main} and Fig.~\ref{fig:g5}. This procedure effectively removes extended
radio emission on angular scales $>0\rlap{.}''7$ while maintaining the angular resolution. 
In the \dr map the effect of cutting the shortest baselines is seen as a 
reduction of the noise level and an improved visual identification 
of CRSs.  
In the \gf map, although the noise was reduced with self-calibration, its level is still high (44\,$\mu$Jy beam$^{-1}$)
and only two CRSs are identified outside the shell.
Future instruments with higher angular resolution, such as the next-generation VLA, will be needed to search for fainter CRSs in G5.89.
%Thus, we focus mainly on \dr\ and present first the results obtained for \dr, followed by those for \gf.

\subsection{Compact radio sources in \dr}\label{sec:CRS-DR21}

The FB image using only long baselines was visually inspected,
and criteria were established to identify the CRSs.
A CRS was defined as a source with an intensity exceeding 5$\sigma$, where $\sigma$
is 30\,$\mu$Jy\,beam$^{-1}$ near the arcs or within the core, and 14\,$\mu$Jy\,beam$^{-1}$
farther away. 
In addition, the source must have a size of at least one synthesized beam and 
must not lie near, or be part of, extended emission.
Following these criteria, 13 CRSs were identified in \dr.
Nine of them fall in the area of the \dr core; see 
Fig~\ref{fig:main}. The source properties (position and flux density)
were obtained using the CASA task {\it imfit}, which performs a 2D 
Gaussian fit to the pixel intensities. The source positions and fluxes 
are listed in Table~\ref{tab:crsM}.

\subsubsection{Variability}

Sources were also identified in the individual epoch maps,
where all sources were detected in the first epoch 
but only eleven in the second. Flux densities and upper limits for
non-detections are listed in Table~\ref{tab:crsM}.
To determine source variability between both epochs, the 
measured flux densities were compared. Specifically, for each source, 
the difference between the flux densities is estimated and normalized 
by the maximum flux. The resulting values, 
expressed in percentage, are also listed in Table~\ref{tab:crsM}. 

\subsubsection{Spectral index}

The source flux density varies across the observed frequency.
This variation is a power-law function with the shape 
\begin{equation}\label{eq:1}
    S_\nu\propto\nu^\alpha
\end{equation}
\noindent where $\alpha$
is the spectral index. To obtain the spectral index of the 
detected CRS, we measured their fluxes in the 
LSB and USB images, both in the individual epochs and in the combined
epochs. The measured flux densities are used to estimate the spectral 
index by using Equation~\ref{eq:1}. The spectral indices obtained
for the combined epoch images are listed in Table~\ref{tab:crsM},
while those for the individual epoch images, they are given Table~\ref{tab:crsA} 
in Appendix~\ref{app:additional-properties}.

\begin{table*}
\setlength{\tabcolsep}{3.5pt}
\begin{center}
\small
%\scriptsize
 \caption{Compact radio sources in \dr.}
    \begin{tabular}{ccccccccccc}
         \hline
         \hline
\# & R.A. & Dec. & $S_{\nu,{\rm int}}$\supa& $S_{\nu,{\rm peak}}$\supa &$\alpha$ & 
$S_{\nu,{\rm int,1}}$& $S_{\nu,{\rm peak,1}}$ & 
$S_{\nu,{\rm int,2}}$& $S_{\nu,{\rm peak,2}}$ & Var. \\
 &$20^{\rm h}39^{\rm m}[^{\rm s}]$&$42^{\circ}19'['']$&
 ($\mu$Jy) & ($\mu$Jy bm$^{-1}$)& &
($\mu$Jy) & ($\mu$Jy bm$^{-1}$)& 
($\mu$Jy) & ($\mu$Jy bm$^{-1}$)&  (\%)\\
 \hline
1& 0.042& 36.48& $235\pm 19$& $223\pm 10$& $0.5\pm 0.5$& $202\pm 14$& $201\pm 7$& $290\pm 27$& $282\pm 15$& $30\pm 8$ \\
2& 0.369& 30.62& $138\pm 27$& $109\pm 13$& $1.2\pm 2.0$& $178\pm 29$& $179\pm 17$& ...& $<54$& $>69\pm12$\\
3& 0.437& 58.69& $133\pm 22$& $109\pm 11$& $0.3\pm 1.5$& $148\pm 24$& $136\pm 13$& $98\pm 26$& $95\pm 14$& $33\pm 18$\\
%4& 0.554& 46.16& $115\pm 15$& $93\pm 7$& $-0.2\pm 1.6$&  $49\pm 25$& $51\pm 15$& $152\pm 18$& $146\pm 10$& $67\pm12$\\
%5& 0.848& 37.04& $261\pm 44$& $142\pm 16$& $-0.4\pm 1.8$& $201\pm 59$& $131\pm 25$& $210\pm 47$& $155\pm 21$& $4\pm 17$\\
4& 0.869& 38.89& $315\pm 73$& $176\pm 28$& $-0.3\pm 1.6$& $322\pm 71$& $183\pm 27$& $300\pm 93$& $178\pm 37$& $6\pm 18$\\
%7& 0.904& 31.60& $123\pm 31$& $93\pm 14$& $1.7\pm 2.7$& $143\pm 72$& $73\pm 26$& $181\pm 36$& $149\pm 18$& $20\pm 26$\\
5& 1.102& 35.26& $371\pm 45$& $223\pm 18$& $-0.8\pm 1.0$& $351\pm 56$& $223\pm 23$& $371\pm 55$& $227\pm 22$& $5\pm 11$\\
6& 1.165& 38.50& $9770\pm 620$& $2120\pm 110$& $-0.2\pm 0.4$& $9750\pm 640$& $2450\pm 130$& $9870\pm 600$& $2340\pm 120$& $1\pm 5$\\
7& 1.190& 18.64& $107\pm 15$& $97\pm 7$& $-0.6\pm 1.3$& $145\pm 19$& $142\pm 11$& ...& $<48$& $>66\pm12$\\
8& 1.215& 35.51& $385\pm 32$& $339\pm 16$& $-0.4\pm 0.7$& $403\pm 40$& $368\pm 21$& $379\pm 51$& $347\pm 27$& $5\pm 9$\\
9& 1.234& 37.37& $357\pm 38$& $211\pm 15$& $0.7\pm 1.0$& $300\pm 52$& $204\pm 23$& $356\pm 39$& $227\pm 17$& $15\pm 12$\\
10& 1.329& 34.33& $657\pm 47$& $535\pm 23$& $0.5\pm 0.6$& $641\pm 43$& $548\pm 22$& $600\pm 52$& $549\pm 28$& $6\pm 7$\\
%14& 1.403& 40.31& $462\pm 115$& $123\pm 25$& $-4.4\pm 2.6$& $307\pm 92$& $146\pm 31$& $266\pm 80$& $112\pm 24$& $13\pm 19$\\
11& 1.475& 34.61& $1860\pm 120$& $937\pm 41$& $0.2\pm 0.5$& $1514\pm 91$& $945\pm 37$& $1710\pm 110$& $981\pm 44$& $11\pm 7$\\
%16& 1.712& 38.07& $155\pm 33$& $103\pm 14$& $-3.6\pm 2.5$& $216\pm 50$& $120\pm 18$& $219\pm 38$& $113\pm 13$& $1\pm 14$\\
12& 1.733& 36.40& $782\pm 56$& $364\pm 19$& $0.2\pm 0.7$& $571\pm 54$& $361\pm 23$& $695\pm 66$& $377\pm 25$& $17\pm 10$\\
13& 1.825& 39.01& $505\pm 49$& $232\pm 16$& $0.7\pm 1.0$& $477\pm 82$& $213\pm 26$& $374\pm 49$& $236\pm 21$& $21\pm 13$\\
         \hline
    \end{tabular} \label{tab:crsM}\\
\end{center}
\begin{list}{}{}
    \item[\supa] Fluxes measured in the combination of the two epochs.
\end{list}
\end{table*}%\end{landscape}

\subsubsection{Circular polarization}

The Stokes V images were inspected at the position of detected radio sources. In the images of the individual epochs no source was detected. In the image obtained by combining the two epochs, only source \#3 is detected at a 3$\sigma_{\rm rms}$ of $33\pm11\mu$Jy, indicating a right handed circular polarization of $26\pm11$\%. 

As the noise level of the Stokes V image (given in 
Table~\ref{tab:obs}) is uniform, it can be stated that the upper limit of source detections 
in these images are 36, 45 and  33\,$\mu$Jy beam$^{-1}$, for the first, second, and combined epochs, respectively.  

\subsubsection{Counterparts}

A search for counterparts of the CRSs was performed using
of the SIMBAD astronomical database\footnote{\url{https://simbad.cds.unistra.fr}}.
The search was constrained to a radius of $1''$. Sources 1 and 2 were found to have 
counterparts in X-rays and infrared (IR), and sources 3 and 7 were found to have 
counterparts in X-rays. Three sources have been identified as young 
stellar objects (YSOs). These results are listed in Table~\ref{tab:count}. 
In this table, we also provide the radio luminosities 
estimated using the equation 
$$L_R[{\rm erg\,s}^{-1}\,{\rm Hz}^{-1}]=1.19\times10^{15}\left(\frac{S_\nu}{\mu{\rm Jy}}\right)\left(\frac{D}{\rm kpc}\right)^2$$
\noindent where $S_\nu$ is the total flux density and $D$ the distance to the radio source.
The remaining sources had no counterparts at any other wavelength.

\begin{table*}%[h!]
\small
\def\arraystretch{1.2}
    \begin{center}
    \caption{Counterparts to compact radio sources of \dr. }
    \label{tab:count}
    \begin{tabular}{cccccc|c}
    \hline\hline
    Source \# & Infrared & X-rays & $L_X$ [erg s$^{-1}$]\supa & Type & Ref.&$L_R$\supb [erg s$^{-1}$\,Hz$^{-1}$]\\
    \hline
   % \#     &
1   & 2MASS J20390005+4219364 & MPCM J203900.04+421936.5 & $1.08\times10^{31}$ &YSO& \citet{broos2013}& $6.3\times10^{17}$\\
2   & UGPS J203900.35+421931.7 & MPCM J203900.37+421930.8& $1.36\times10^{30}$ &YSO & \citet{broos2013}& $3.7\times10^{17}$\\
3   & ... &  	MPCM J203900.44+421958.8 &$5.31\times10^{30}$ &YSO&\citet{broos2013}&$3.6\times10^{17}$\\
7  & ... & [RJM2014] 15   &$7.08\times10^{30}$&... &\citet{rivilla2014}&$2.9\times10^{17}$\\
         \hline
    \end{tabular}\\
\end{center}
\begin{list}{}{}
    \item[\supa] $L_X$ is the X-rays luminosity.
    \item[\supb] $L_R$ is the radio luminosity.
\end{list} 

\end{table*}

\subsection{The arc-shaped structures in \dr}
\label{sec:arc-shap-struct-results}
The high angular resolution images allow us to resolve and appreciate
details in \dr that have not been seen before. \citet{harris1973} initially 
identified that this \ion{H}{II} region can be resolved into four compact condensations. 
Three of them are related to the \dr core,  one at the west, north, and south. 
The fourth condensation, is located $\sim 20''$ to the north of the \dr core, as seen in Fig.~\ref{fig:main}. 
Though these have been observed with better angular resolution than \citet{harris1973} observations
\citep[e.g.,][]{cyganowski2003}, in our new radio image, the classical three-condensation 
morphology of the \dr core is decomposed into a series of filamentary and arc-shaped
bright rims. 

Such features are commonly associated with density gradients in 
the parent molecular cloud, causing an irregular expansion of the 
\ion{H}{II} region 
\citep[e.g.,][]{garcia1996,comeron1997,arthur2006}. To our knowledge,
this kind of arc-shaped structures has not previously been reported 
at radio wavelengths in other compact \ion{H}{II} regions, although similar 
structures have been observed at radio frequencies in more evolved 
\ion{H}{II} regions such as, for example, the Orion Nebula 
\citep{forbrich2016} and LBN 978 \cite{carrasco2006}. In those
cases, the apex of the curved bright rims typically points toward the 
massive star that is photo-ionizing them. 

Our findings suggest that the observed arcs in \dr may also trace 
ionization fronts or shock interfaces, indicating active shaping 
of the nebula by its embedded massive stars. We can use their morphologies
to constrain the location of the massive stars that ionize them.

\subsubsection{Parabola fit} \label{sec:parabola}

In order to characterize the arc-shaped structures observed in our \dr radio
images, we fit them with a parabolic shape. For this analysis, 
we have used the FB image containing all the baselines and combined the two epochs, 
as this image is the most sensitive to the extended emission. 

\begin{figure*}%[!ht]
 \centering
\begin{tabular}{ccc}
\includegraphics[height=0.14\textheight]{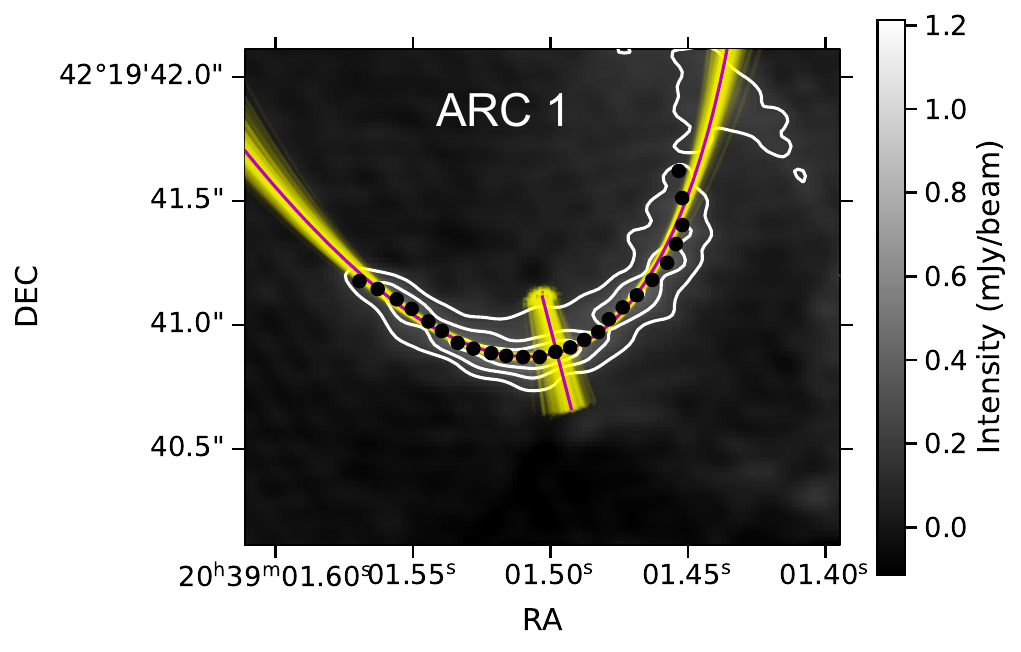}&
\includegraphics[height=0.14\textheight]{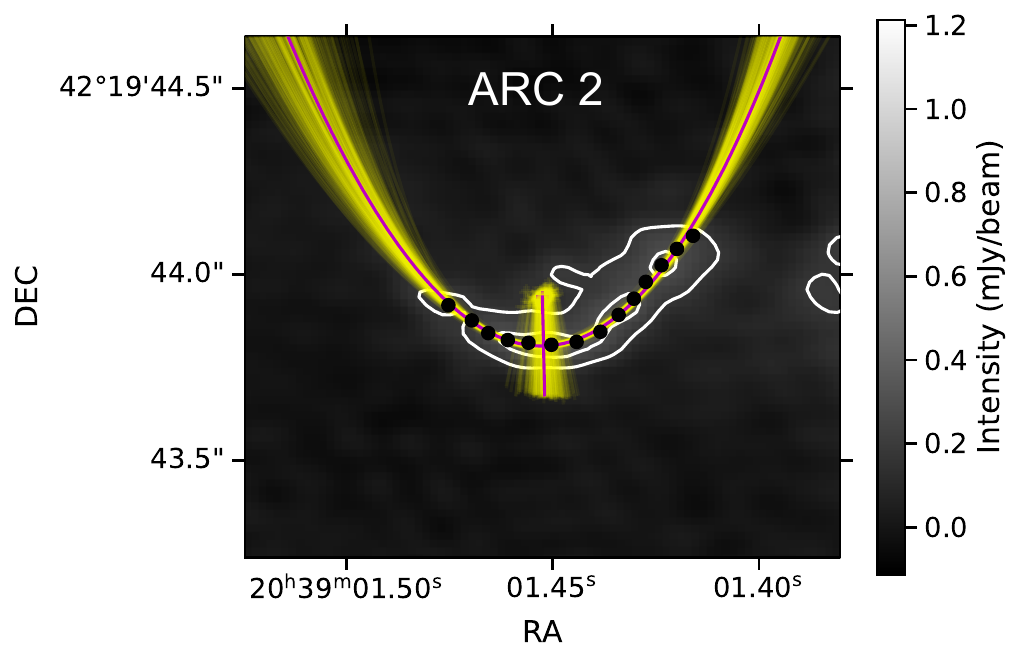}&
\includegraphics[height=0.14\textheight]{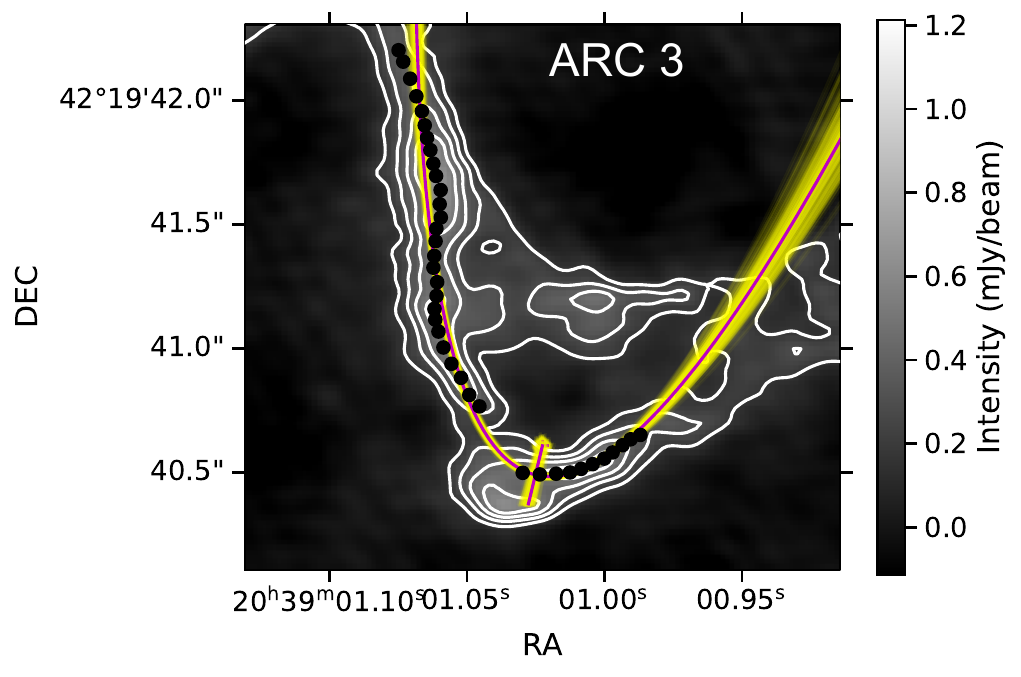}\\
\includegraphics[height=0.14\textheight]{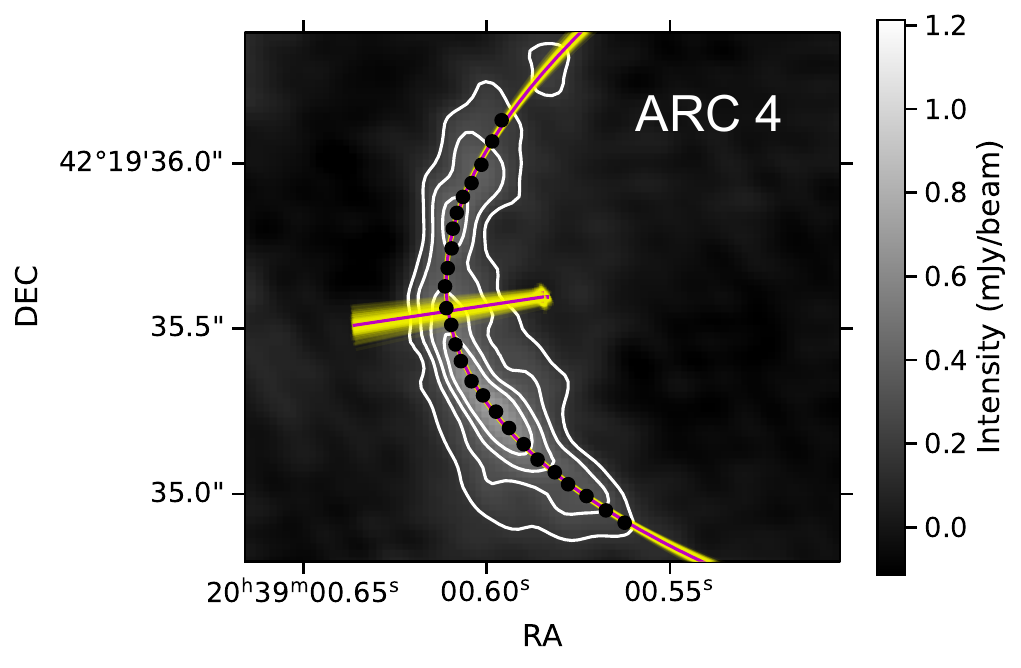} &
\includegraphics[height=0.14\textheight]{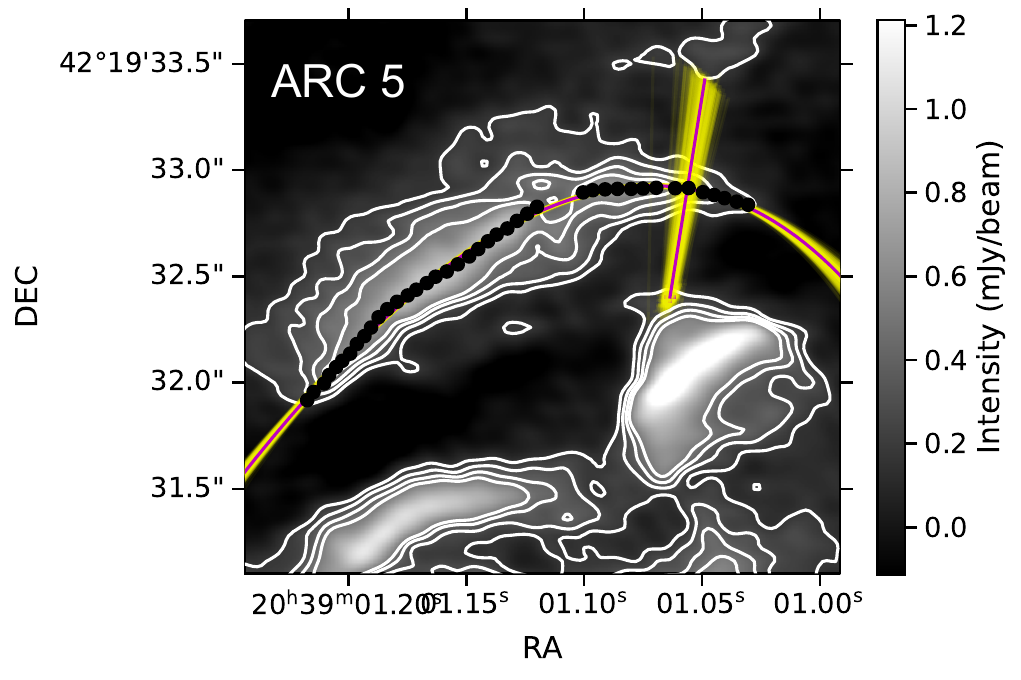}&
\includegraphics[height=0.14\textheight]{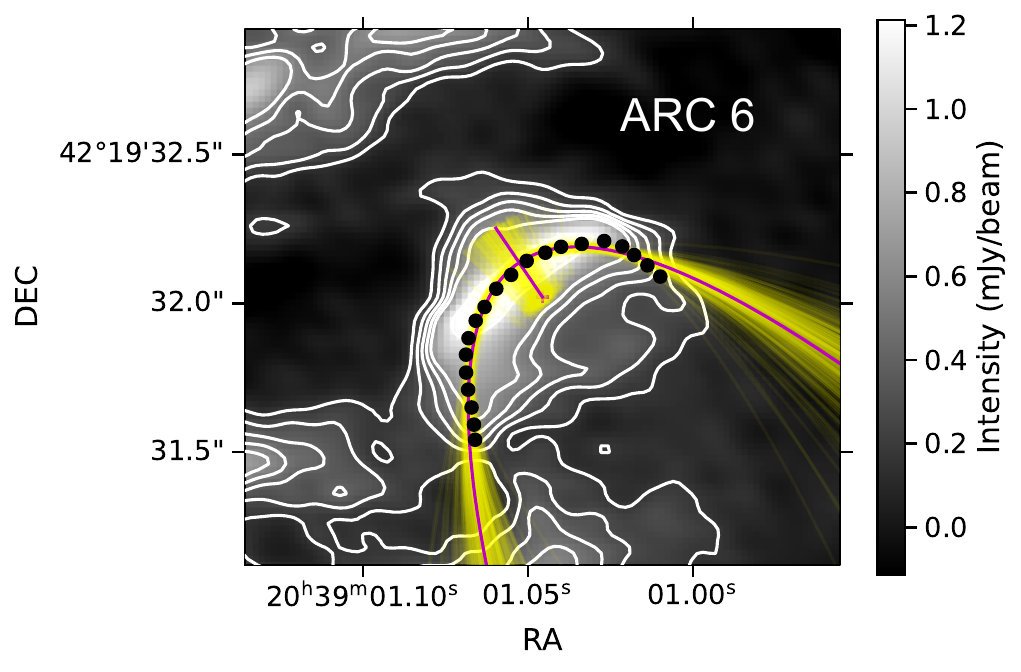}\\
\includegraphics[height=0.14\textheight]{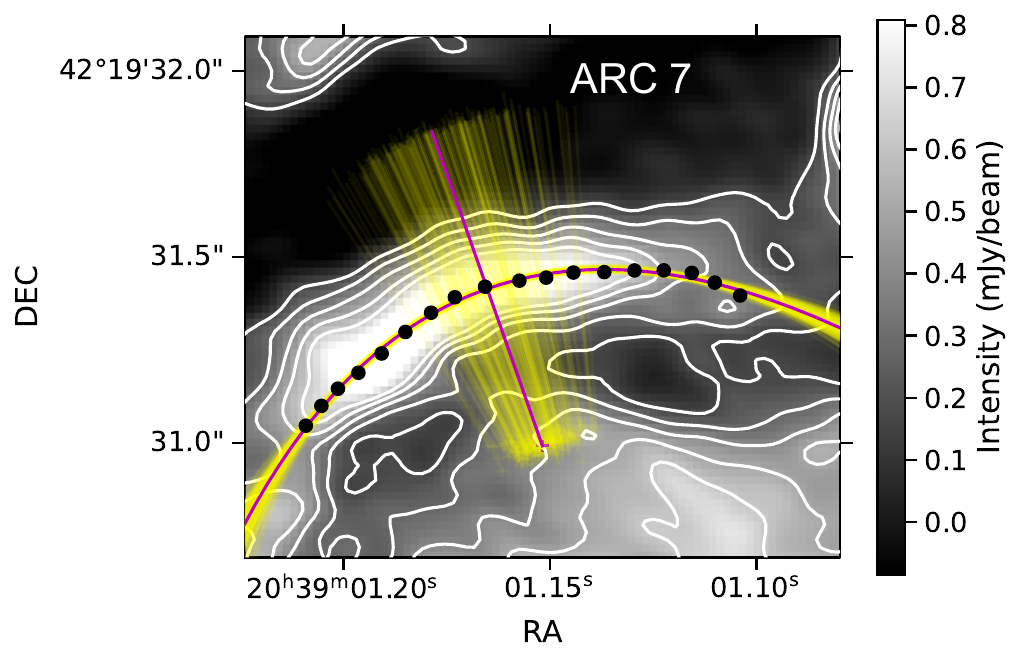}&
\includegraphics[height=0.14\textheight]{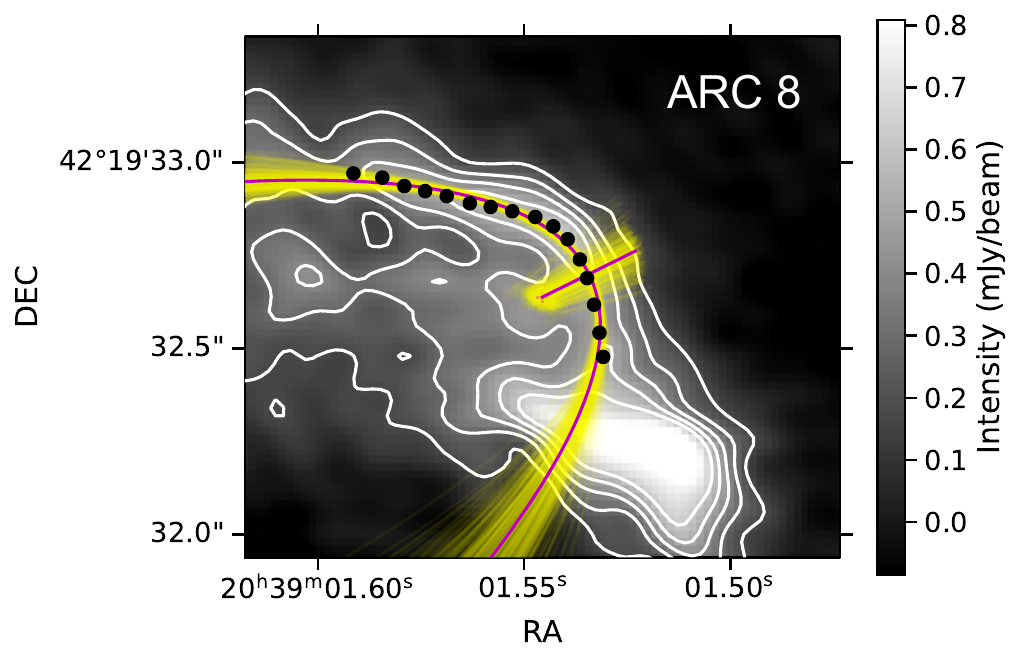}&
\includegraphics[height=0.14\textheight]{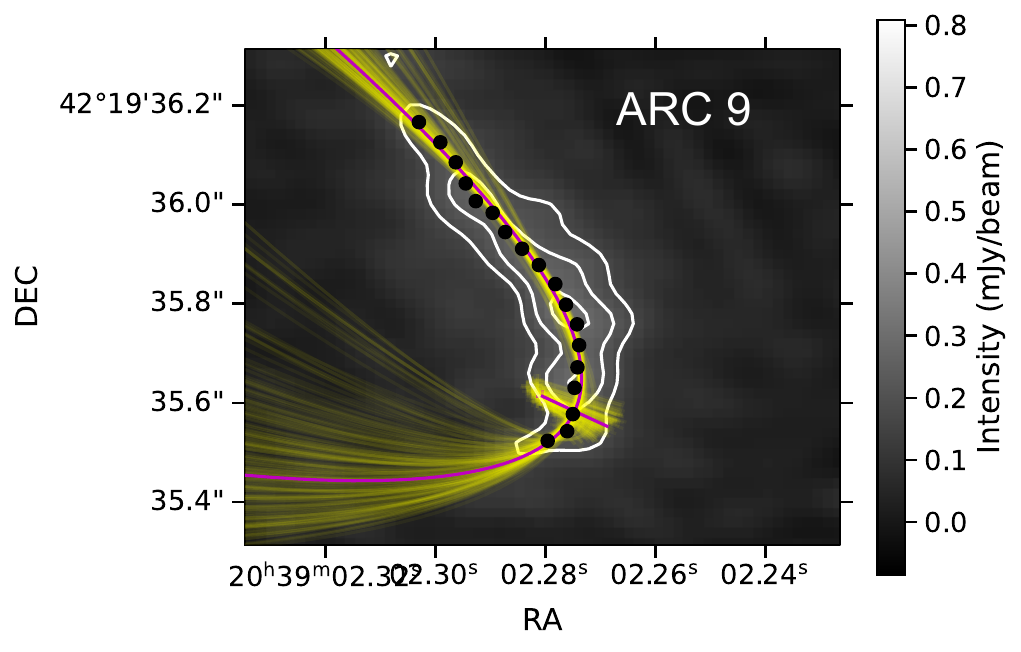}
\end{tabular}
 \caption{Detailed view of the parabola fit for each arc in DR21 cm image in order from 1 to 9. The black points are the data taken from the observations. The magenta parabola shows the best fit for each arc, the magenta line reveals the orientation of the parabola, and the magenta cross presents the parabola focus $x_f$, $y_f$. The yellow lines indicate the different possibilities found using MCMC ensemble sampling. Contours are 3, 5, 7, 9, 11, 15 times 56$~\mu$Jy~beam$^{-1}$.}
 \label{fig:arcs}
\end{figure*}

The process consists of two steps. 
First, the arc is traced with a series of points  
manually defined following the trace of the maximum intensity contours until reaching 3-$\sigma$ or finding other structures. Second, the selected points are fitted using a custom Python routine \footnote{Developed by W. Henney; the code is available at \url{https://github.com/div-B-equals-0/confitti}.} to derive a parabola.
This routine is described in Appendix \ref{app:pap} and employs a two-step approach: first, it performs a numerical minimization to derive the best-fit curve. Then, it carries out a Markov Chain Monte Carlo (MCMC) analysis to derive the uncertainties in the curve parameters. The best-fit parameters derived for each arc are presented in Table~\ref{tab:par_fit}, and a detailed view of the fit is shown in Figure~\ref{fig:arcs}.

Figure \ref{fig:arcs} shows that the parabolic fits align well with most arcs except for arcs 5 and 9. Specifically, Arc 5 appears to be a superposition of two arcs, which may cause inconsistencies in the fit compared to the others. In the case of the Arc 9, the missing sensitivity of one of the tails of the arc may cause a problem in the fit to align the parabola in the direction of the maximum intensity peak.

\begin{table*}%[!ht]
    \centering
      \caption{Arc-shaped structure flux densities and parameters of the best parabola fit for each arc of \dr.}
    \begin{tabular}{c c c c c c c }
    \hline
    \hline
        ID & $S_{\nu, \rm int}$\supa& $S_{\nu,{\rm peak}}$\supa  & $x_f$\supb & $y_f$\supb & $r_0$\supc & $\theta$ \supd  \\
           & mJy & mJy beam$^{-1}$& $20^{\rm h}39^{\rm m}[^{\rm s}]$&$42^{\circ}19'['']$& [$''$] & [$^\circ$] \\
    \hline
Arc 1 & 11.3 & 0.52 $\pm$ 0.81 & $ 1.5036\pm 0.0015$ & $41.104\pm 0.012$ & $ 0.230\pm 0.006$ & $194\pm3$ \\ 
Arc 2 & 2.8 & 0.32 $\pm$ 0.23 & $ 1.4535\pm 0.0009$ & $43.924\pm 0.010$ & $ 0.130\pm 0.006$ & $181\pm3$ \\ 
Arc 3 & 26.2 & 0.66 $\pm$ 0.37 & $ 1.0235\pm 0.0007$ & $40.598\pm 0.012$ & $ 0.120\pm 0.006$ & $166\pm1$ \\ 
Arc 4 & 13.46 & 0.71 $\pm$ 0.33 & $ 0.5862\pm 0.0004$ & $35.586\pm 0.008$ & $ 0.288\pm 0.004$ & $98\pm1$ \\ 
Arc 5 & 51.58 & 11.35 $\pm$ 0.43 & $ 1.0645\pm 0.0012$ & $32.392\pm 0.032$ & $ 0.518\pm 0.032$ & $-9\pm2$ \\ 
Arc 6 & 28.15 & 1.49 $\pm$ 0.79 & $ 1.0464\pm 0.0016$ & $32.012\pm 0.024$ & $ 0.138\pm 0.008$ & $33\pm5$ \\ 
Arc 7 & 20.03 & 1.09 $\pm$ 0.67 & $ 1.1530\pm 0.0040$ & $30.984\pm 0.024$ & $ 0.444\pm 0.018$ & $19\pm6$ \\ 
Arc 8 & 6.0 & 0.6 $\pm$ 0.43 & $ 1.5464\pm 0.0015$ & $32.626\pm 0.014$ & $ 0.138\pm 0.016$ & $-64\pm5$ \\ 
Arc 9 & 2.5 & 0.3 $\pm$ 0.22 & $ 2.2820\pm 0.0013$ & $35.590\pm 0.020$ & $ 0.066\pm 0.020$ & $244\pm7$ \\ 
\hline
    \end{tabular}\label{tab:par_fit}\\
\small
    \def\arraystretch{0.5}
    \begin{list}{}{}
        \item[\supa] Fluxes measured in the image combining the two epochs.
        \item[\supb] $x_f$ and $y_f$ are the positions of the focus in R.A. and Dec. (J2000) respectively.
        \item[\supc] $r_0$ is the distance between the focus and the parabola apex.
        \item[\supd] $\theta$ is the angle given from the north to the east. 
    \end{list}
\end{table*}

\subsection{Compact Radio Sources in \gf}

The extended and bright emission of the \gf\ shell makes the identification of CRSs in this area a difficult task, even though the image with baselines larger than 300 k$\lambda$ was used. No clear compact sources fulfilling the criteria outlined in Sec.~\ref{sec:CRS-DR21} were found near the shell. Outside the shell, the same criteria were applied using the image with baselines $>300$ k$\lambda$ and a local rms of 20 $\mu$Jy beam$^{-1}$. Following this, two CRSs were detected and are marked in the left panel of Fig.~\ref{fig:g5}. Their properties, measured in the combined-epoch images with $uv$-range $>300$ k$\lambda$, were obtained using the CASA task {\it imfit}. The source positions and fluxes are shown in Table \ref{tab:crsGf}. Future instruments with higher angular resolution, dynamic range, and sensitivity, such as the next-generation VLA, could reveal additional CRSs in \gf\ for further study.

\begin{table*}
\setlength{\tabcolsep}{3.5pt}
\begin{center}
\small
%\scriptsize
 \caption{Compact radio sources in \gf.}
    \begin{tabular}{ccclccllc}
         \hline
         \hline
\# & R.A. & Dec.  &$S_{\rm peak}$\supa& $S_{\rm int}$& $S_{\rm peak}$&$S_{{\rm peak,LSB}}$&$S_{{\rm peak,USB}}$&$\alpha$\supb \\
%&$S_{\nu,{\rm int,1}}$& $S_{\nu,{\rm peak,1}}$ & 
%$S_{\nu,{\rm int,2}}$& $S_{\nu,{\rm peak,2}}$ & Var. \\
 &$18^{\rm h}00^{\rm m}[^{\rm s}]$&$-24^{\circ}04'['']$ &($\mu$Jy bm$^{-1}$) &
 ($\mu$Jy) & ($\mu$Jy bm$^{-1}$) &($\mu$Jy bm$^{-1}$)  &($\mu$Jy bm$^{-1}$) & \\
% & ($\mu$Jy) & ($\mu$Jy bm$^{-1}$)& 
%($\mu$Jy) & ($\mu$Jy bm$^{-1}$)&  (\%)\\
 \hline
1& 28.109& 55.34  &$547\pm 56$& 408$\pm 13$& $390\pm 13$&$437\pm 20$&$345\pm 13$& $-1.2\pm 0.3$ \\
2& 30.843& 37.82 &$225\pm 15$& $192\pm 10$& $192\pm 10$&$186\pm 12$&$186\pm 14$& $0.0\pm 0.5$\\

         \hline
    \end{tabular} \label{tab:crsGf}\\
\end{center}
Note: Unless otherwise stated, fluxes were measured from the images obtained by combining the two epochs with a $>$300\,k$\lambda$ baseline cutoff.
$^{\rm a}$ Intensity measured in the image containing all baselines.
$^{\rm b}$ Spectral index measured using the fluxes of LSB and USB images 
\end{table*}

\subsubsection{Spectral Index}

Spectral index of the two CRSs identified in \gf 
were determined following equation \ref{eq:1}, 
using the fluxes measured in the LSB and USB images obtained for baselines
above 300\,k$\lambda$ and combining the two epochs. The details of the fluxes
for each image is also presented in Table \ref{tab:crsGf}. From the table,
source \#1 shows a negative spectral index, however its nature needs to be further 
investigated.

\subsubsection{Counterparts}

A search for counterparts of the CRSs in \gf is also done with the help
of the SIMBAD astronomical database with the same radius used in the \dr search of $1''$ . No counterparts were found using that radius.
The nearest counterparts to the CRSs lie more than $10''$ away and are reported as young stellar object candidates by the Spitzer/IRAC Candidate YSO Catalog \citep[SPICY,][]{Kuhn2021}.
The nearest objects are 2MASS J18002733-2404427 for \#1 and  2MASS J18003158-2404389 for \#2, located at a distance of $16.36''$ and $10.78''$ respectively from our observed positions. 

\subsection{Extended emission in \gf}

As shown in Figure~\ref{fig:streams}, the shell-like structure of \gf\ resembles 
that found in NGC~6334A, where the most recent interpretation of its central
source suggests a dynamical scenario in the past that produces runaway stars 
and an ionizing shell with a similar square-like shape \citep{yanza2025}. 
In contrast, no clear arcs are visible within the shell of \gf, unlike in \dr.
Nevertheless, a parabolic fit was applied to a possible arc along the southern 
part of the shell, as presented in Appendix~\ref{app:gf}.

\section{Discussion}

Most of the detected CRSs lie well inside the \dr core, thus 
they are likely related to young stars. By using equation A11
of \citet{anglada1998}, we estimated that the number of expected extragalactic 
background sources is 0.04; limited to the imaged area, at a central frequency 
of 15\,GHz and with a minimum flux density value of 60\,$\mu$Jy. The low
number of expected background radio sources aligns with the idea that all detected CRSs are young stars. 

%possible relation to the explosion event
Given the limitation of our observations it is not possible to determine
the proper motions of the detected CRSs. However, sources 4, 5, 6, 8, 9, 10, and 11, are located within or close to the areas of 
the suggested origin of the \dr \exofs, see Fig.~\ref{fig:main}, and this make them good 
targets for future astrometric follow-ups. 

\subsection{Radio emission nature of compact radio sources}

CRSs are common in star-forming regions \citep{rodriguez2012}.
These are usually related to young stars, with scarce
sources being background extragalactic radio sources. 
\citet{rodriguez2012} reviewed the properties at radio frequencies
of CRSs, we point the exhaustive reader to this work and references therein 
for a more detailed discussion, that we now summarize.
Radio emission processes can be thermal and nonthermal. 
Thermal radio emission is produced by thermal electrons that are 
accelerated in the vicinity of ions, while nonthermal radio 
emission is produced by electrons moving at relativistic speeds in
magnetic fields. The compact thermal radio emission is found to be 
related to the base of jets of YSOs, externally photoevaporated 
disks (proplyds), knots of jets, and hyper-compact \ion{H}{II} regions. 
On the other hand, compact nonthermal radio emission can be related to
knots of jets of massive stars, colliding wind regions, and 
magnetically active low-mass stars. 

The measured radio properties can help distinguish between 
the emission processes at play and thus give clues about the 
nature of the radio sources \citep[see Table 1 by][]{rodriguez2012}. 
Thermal radio sources are expected to have $\alpha>-0.1$,
no significant variability, and no polarization. On the contrary, 
nonthermal radio sources can be divided into two sub-groups, depending on if they are
associated with massive stars or with low mass stars. 
Nonthermal radio sources associated with massive stars are related to strong shocks,
either due to jet interaction with the interstellar medium \citep{carrasco2006}, colliding wind regions \citep{dzib2013,yanza2025}
or massive runaway stars \citep{yanza2025}. These nonthermal radio sources have negative spectral indices,
low variability and can show linear polarization.  On the other hand, magnetically active low-mass stars
usually have negative spectral indices, but they can have values from $-2$ to 2 \citep{dulk1985}. The radio emission from
these source show variations $>50$\% in a few days, and can show circular or linear polarized emission \citep[e.g.,]{rodriguez2012,dzib2015}. 

The spectral index is a good discriminator between thermal and 
nonthermal processes. However, the large errors in the measured
spectral indices from our observations prevent their use for this purpose. 
Circular polarization is also a good discriminator, and we found 
source \#3 with a circular polarization of $26\pm11$\,\%. This 
source also shows a variability of $33\pm18$\,\%, and it has a 
counterpart at X-rays and is classified as YSO by \citet{broos2013}. 
This could indicate that this compact radio source is related to 
a magnetically active young star. Radio sources \#2 and \#7
have variability above 60\%, and 
are also X-ray emitters.
These properties point towards these sources being 
magnetically active stars; however, as their spectral indices are not 
well constrained, this needs further investigation. The remaining compact radio 
sources show variability $\lesssim30$\%, and are likely  to be
thermal radio emitters; however, the phenomena producing them 
cannot be identified with the current data.

\subsection{Comparison with CRSs in other star forming regions}

CRSs have been detected in other
star-forming regions. The Orion Nebula Cluster (ONC)
is an outstanding case with about 600 detected CRS
\citep{forbrich2016,vargas2021}, including those
related to the BN/KL \exofs.  Other regions include
M\,17 with 182 CRSs \citep{yanza2022} and NGC\,6334 D-F 
with 83 CRSs \citep{medina2018}. 

Although the ONC and \dr are at different stages of star 
formation and activity, meaningful comparisons can be made. 
To simplify our comparison, we assume that the source count is independent of the spectral frequency at which they are observed. 
This assumption is supported by the results of \citet{forbrich2016}, 
who measured the spectral indices of 170 CRS. The mean of these is 
$\bar{\alpha}_{\rm ONC-CRS}=0.05\pm0.04$ with a standard deviation 
of 0.54, suggesting that frequency dependence is not dominant.

The weakest sources in \dr have  peak intensities of 
$93\,\mu$Jy beam$^{-1}$, which would correspond to 
1.2\,mJy\,beam$^{-1}$ if placed at the distance of the ONC.
We also notice that in the \dr core, 14 CRSs are detected within 
a radius of $10''$. For comparison, the ONC contains 31 CRSs 
with peak intensities above 1.2\,mJy\,beam$^{-1}$.
The densest region of CRSs in the ONC is the Orion Trapezium, 
where 21 CRSs are found within a radius of $36''$. This area 
is similar in size to the projected area of the \dr core if 
it were at the ONC distance. Thus, \dr and the ONC exhibit 
comparable CRS densities in their most active regions.

%This similarity in CRS richness between \dr and the ONC suggests that both regions host comparable levels of compact radio emission, despite differences in evolutionary stage. This result highlights the utility of CRS as a diagnostic tool for probing the environments of star formation across different phases.

This similarity in CRS richness between \dr and the ONC suggests that the parental clouds have similar masses. A strong correlation between the number of sources and their parental cloud mass has been reported in previous studies at different wavelengths \citep[e.g.,][]{lada2010, palau2021, Palau2024}, and has been theoretically explained by several authors \citep[e.g.,][]{ballesteros2024} and numerically demonstrated by \citet{ZamoraAviles2025}. 
This result highlights the utility of using the surface density of CRSs as a diagnostic tool to study the clusters associated with massive star-forming regions.

\subsection{The arc-shaped structures}
\label{sec:arc-shap-struct}

The arc-shaped structures seen in our images are likely produced by the expansion of the \ion{H}{II} region. As described above, we fit each arc with a parabola. Using the resulting parabola parameters, we can find the approximate location of the ionizing massive star (or stars) based on the supposition that their apexes point toward them. 

\begin{figure*}
    \centering
    \includegraphics[height=0.65\textheight]{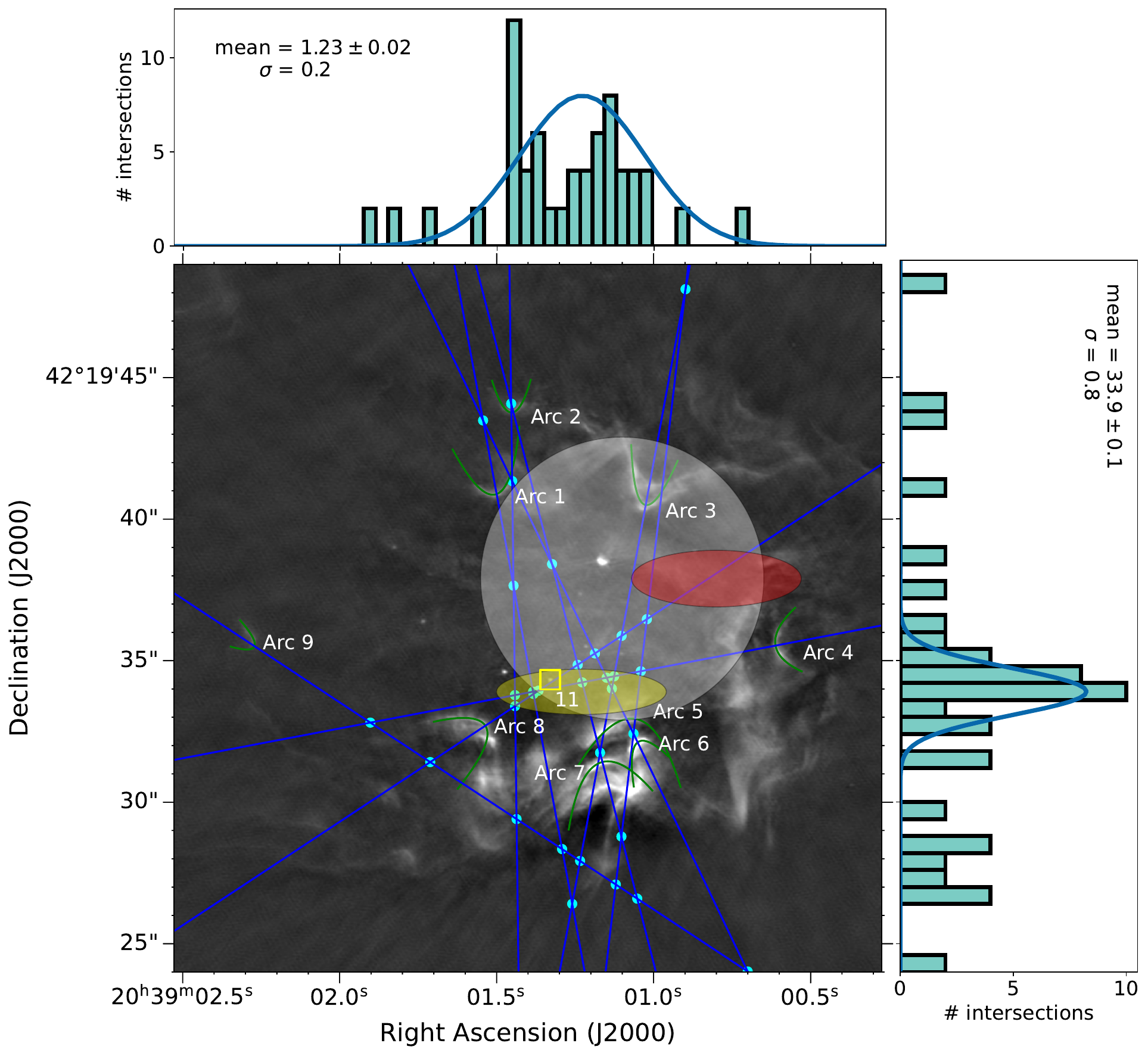}
    \caption{Projection of the symmetry axis of each parabola fitted to each of the identified arcs in DR21. Cyan dots indicate the intersection point of each pair of lines. The yellow transparent ellipse is centered in the mean position of the line intersection points, and the size reflects the standard deviation. The white circle shows the central position of the explosion reported by Zapata et al. (2013), with its size corresponding to the uncertainties. Similarly, the red ellipse represents the central position of the explosion recently suggested by \citet{guzman2024}.  {\it Upper and right panel} are the distribution histograms of 
    the intersection points in the R.A. and Declination coordinates, respectively.}
    \label{fig:all_lines}
\end{figure*}

First, for each arc, we calculated the line passing through the parabola focus and apex, extending it to the center of \dr (see Figure \ref{fig:all_lines}). Second, we determined the intersection point for every pair of these lines, yielding a total of 36 intersections for the nine arcs. The intersection points are shown in the main panel of Figure~\ref{fig:all_lines}, while the right ascension and declination distributions are shown in the top and right panels, respectively. Fitting these distributions with a Gaussian, we find the mean and standard deviation, naturally discarding outlier points.

The mean of the intersection positions is $\alpha$(J2000) = \rahms{20}{39}{01}{23}
$\pm0\,\rlap{.}^{\rm s}20$ and $\delta$(J2000)
= \decdms{42}{19}{33}{9}$\pm0\rlap{.}''8$. 
Within measured uncertainties, only one CRS,
\#11, lies at this position. This source shows no signs 
of variability or any other sign of nonthermal radio emission, and at any other wavelength 
counterpart has yet been identified. Consequently, we cannot confirm or rule out its role as the primary ionizing source in \dr. Nonetheless, its alignment with the pointing direction of the arc-apexes is remarkable and merits further investigation.

We also compare the parabola intersection position with the suggested centers for the \dr \exof. It is consistent, within uncertainties, with the center proposed
by \citet{zapata2013}, but it is not with the more recently suggested center by \citet{guzman2024}. Because none of the stars have been firmly related to either the explosive event or the source of the ionizing photons maintaining the \dr \ion{H}{II} region, it is unclear if the stars participating in these phenomena are the same or connected. The position inconsistency could indicate that they are likely unrelated. 
However, we note that the position center of the explosion favored by \citet{guzman2024} was obtained from a statistical analysis using only six out of the 18 CO streamers they detected. Thus, a more comprehensive statistical analysis of all CO streamers could provide additional insight.

\subsection{The center of explosion}

\begin{figure*}
    \centering
    \includegraphics[width=0.8\textwidth]{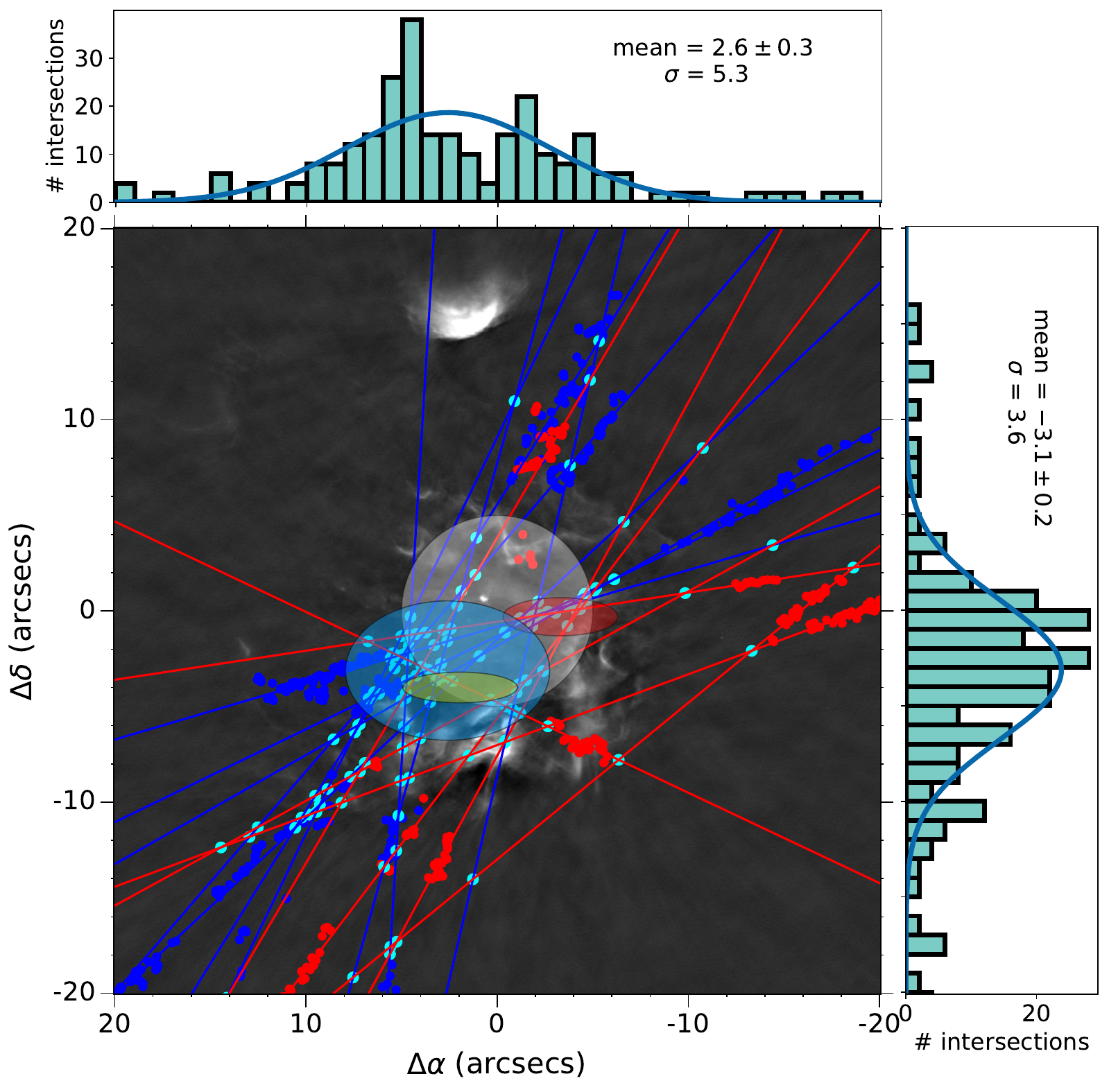}
    \caption{Center of the explosion from the CO streamers. {\it Main panel:} CO streamer superposed on the VLA 15~GHz image. For a direct comparison with \citet{guzman2024}results, the image is shown relative to $\alpha$=\rahms{20}{39}{01}{1} and $\delta=$\decdms{+42}{19}{37}{9}; the first suggested center of the DR\,21 \exofs \citep{zapata2013}. Blue and red dots indicate the blue and redshifted streamers, respectively, from \citet{guzman2024}, and the blue and red lines are the best resulting line fit to each of them. Cyan dots indicate the intersection point of each pair of lines. The light blue transparent ellipse is centered in the mean position of the line intersection points, and the size reflects the standard deviation.
    White, red, and yellow transparent ellipses are  are the same as in Fig.~\ref{fig:all_lines}.  {\it Upper and right panel} are the distribution histograms of 
    the intersection points in the R.A. and Declination coordinates, respectively. }
    \label{fig:CEO}
\end{figure*}

\citet{guzman2024} identified a total of 18 CO outflow
streamers associated with \dr, but used only six of them to determine the explosion center. In top pannel of Figure~\ref{fig:streams}, we show these outflows seen in projection to our VLA image.  In the present work, 
we adopt a similar approach to their analysis while including all 18 streamers.
Each streamer was fitted with a straight line, and every pair of lines was then used to compute their intersection point, yielding a total of $18\cdot(18-1)/2=153$ intersection points. The streamer and the best line fit to them are shown in Figure~\ref{fig:CEO}. 

Following the procedure described for the arc-shaped structures, we fitted the resulting intersection distributions with a Gaussian function, which are shown in Figure~\ref{fig:CEO} .
The mean of the intersections
is found at the position $\alpha$(J2000) = \rahms{20}{39}{01}{27}
$\pm0\,\rlap{.}^{\rm s}35$ and $\delta$(J2000)
= \decdms{42}{19}{34}{8}$\pm3\rlap{.}''6$. 

Notably, this newly derived explosion center closely coincides with the position toward which the arc-shaped structures appear to converge. These findings suggest that the stars shaping the \dr \ion{H}{II} region may also be involved in the \dr \exof, indicating a potential connection between the explosive event and the region ionizing sources.

\subsection{Discussion of individual objects}

\subsubsection{Inner part of \dr.} 
\begin{figure}
    \centering
    \includegraphics[width=1\linewidth]{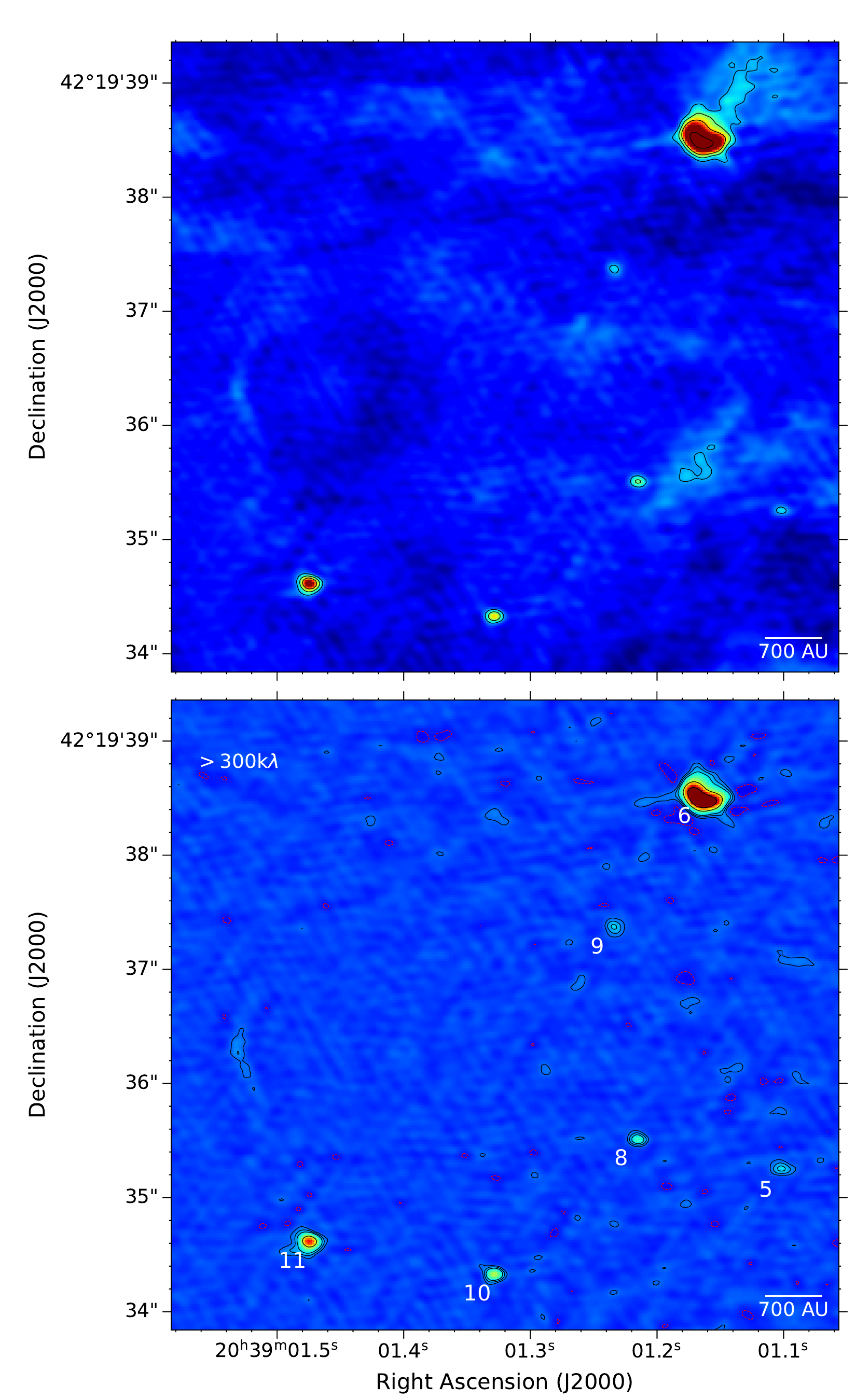}
    \caption{Inner part of \dr\ as seen in the combined maps by using
    the full ($u,\, v$)-range ({\it top panel}) and baselines $>300\,{\rm k}\lambda$ ({\it bottom panel}). Contour levels as in Fig.~\ref{fig:main}. }
    \label{fig:sou9}
\end{figure}

Source \#6 is slightly resolved with a distinct arc-shaped 
structure, see Fig.~\ref{fig:sou9}. 
The recovered structure has a cometary 
shape resembling externally photoevaporated discs
(proplyds) in the ONC
\citep[e.g.,][]{zapata2004}. The apex of proplyds point in the 
direction of the massive star that produces the ionizing photons,
and produce free-free emission with flat or positive spectral index at radio frequencies \citep{rodriguez2012}.
As seen in Table~\ref{tab:crsM}, source \#6 has a relatively flat 
spectrum ($\alpha=-0.2\pm0.4$), and very low variability ($1\pm5$\%)
between observed epochs. These properties suggest that it is a 
thermal radio source. 

The apex of source \#6 points to source \#9 \#10 and \#11, see Fig.~\ref{fig:sou9},
suggesting that one of these sources could externally ionize source \#6.
While the answer to this question needs to be further investigated 
together with determining the nature of those sources. We assume that 
source \#11 is the star ionizing source \#6, and we can estimate the stellar spectral type of the source. 

The angular separation between sources \#6 and \#11 is $4\rlap{.}''55$.
Source \#6 has a deconvolved angular size of 
$(\theta_{\rm maj}\times\theta_{\rm min})$=($0.235\pm0.017'')\times(0.133\pm0.011'')$ with a P.A.=$62\pm6^{\circ}$. 
We assume that from the point of view of source \#9, source \#6 is circular and with a size equal to its major-axis. Thus, the angle
of source \#6 with respect to source \#11 is $\gamma=2\arctan(0.5\cdot\theta_{\rm maj}/\theta_{\rm sep})=2.96^\circ$, or a solid angle of $\Omega = 2\pi \cdot(1 - \cos(\gamma/2.0))= 0.0021$ steradians. Then, source \#6 intercepts 
$\Omega/4\cdot\pi=0.00017$ (0.017\%) of the photons radiated by source \#11. 

The number of hydrogen ionizing photons per second required to keep a nebula ionized \citep[e.g.,][]{martin2003,sanchez2013} can be obtained by
$$N_i=9.0\times10^{43}\left(\frac{S_\nu}{\rm mJy}\right)\left(\frac{D}{\rm kpc}\right)^2\left(\frac{\nu}{\rm 4.9\,GHz}\right)^{0.1},$$
where for source \#6 we have that $S_\nu=9.8$\,mJy, D=1.5\,kpc, and $\nu=15$\,GHz (the mean frequency of our observations). This formulation assumes an environment free of dust and $T_e=10^{4}$\,K. 
Resulting in $N_i=2.2\times10^{45}$\,photons s$^{-1}$, or that source \#11 is producing a total of $2.2\times10^{45}/0.00017=1.3\times10^{49}$\,photons s$^{-1}$, which can be provided by a O7 star \citep[e.g.][]{dzib2013}. This is consistent with the suggested stellar spectral types ionizing the most extended emission of DR\,21 Main \citep[e.g.,][]{roelfsema1989}.

\subsubsection{Sources with X-ray counterparts}

Sources \#1, \#2, \#3, and \#7 have been previously reported as X-ray sources \citep{broos2013,rivilla2014}.  Additionally, \citet{broos2013} identified sources \#1, \#2, and \#3 as YSOs. The high variability of source \#7 also suggests that this is a YSO, as this level of variability is typical of magnetically active low-mass stars \citep[e.g.,][]{dzib2013}.

It is well known that magnetically active stars tend to follow the G\"udel--Benz relation, which links their X-ray and radio luminosities \citep[]{gb1993,gb1994}. These works showed that for main-sequence stars $L_X/L_R\simeq10^{15.5}$\,Hz. However, more recent observations suggest that younger stars have significantly lower values of $L_X/L_R\simeq10^{14\pm1}$\,Hz \citep{dzib2013,dzib2015,yanza2022}. 

Using the luminosity in Table~\ref{tab:count}, we find that the $L_X/L_R$  of these three sources are on the order of $10^{13}$\,Hz. This result suggests that these radio sources are likely magnetically active low-mass stars, consistent with the fact that sources \#2, \#3, and \#7 show signs of nonthermal radio emission. 

%--------------------------------------------------- 
\section{Conclusions}

The VLA high angular resolution observations of 
\dr and G5.89 have unveiled the complex structures of 
these \ion{H}{II} regions, revealing the 
first identification of CRSs in the inner core of DR21 and in the outer parts of G5.89.
We detected a total of 13 CRSs in \dr, 9 of which lie well 
within the core, an area associated with the proposed origin of the explosive molecular outflow (\exofs). Follow-up observations will be crucial to determine whether any of these CRSs are runaway stars connected to the explosive event.
This would support the hypothesis that dynamical disintegration  of multiple massive stellar systems is the origin of
\exofs \citep{bally2011}. Moreover, two CRSs were identified in G5.89, and future telescopes will be essential for detecting fainter sources, especially those located closer to the bright shell structure.

The CRSs detected in our observations in \dr were characterized. 
The properties of three of them indicate they are nonthermal radio 
emitters, suggesting magnetic activity. For most of the remaining sources, the emission nature is 
uncertain and requires further study. We found that source \#6
has a morphology that resembles a proplyd and could be ionized by 
source \#11. We estimate that in this scenario, source \#11 is a 
massive star with a spectral type about O7. The surface number density of CRSs in \dr\ appears comparable to the surface number density in the Orion Nebula Cluster, indicating that the corresponding parental clouds might share similar properties.

We also identify multiple arc-shaped features related to \dr core. Fitting each arc with a parabola, we trace back the pointing direction of their apexes. The results suggest that the arc-shaped structure pointing direction converges near source \#11, which exhibits no signs of variability or nonthermal emission and lacks a confirmed counterpart at other wavelengths. The positional coincidence raises the possibility that this source is a key ionizing star shaping the \dr\ \ion{H}{II} region.

We have also analyzed all 18 outflow streamers previously identified toward \dr\  \citep{guzman2024} to refine the center of the explosive event. We found a new explosion center that coincides, within uncertainties, with the position indicated by the arc apexes. The alignment of the arcs and the outflow streamers on a common center further suggests that the star(s) driving the expansion of the \dr\ \ion{H}{II} region may also be responsible for the \dr \exof.

Overall, our findings highlight how high-resolution radio observations can reveal the intricate interplay of ionization fronts, stellar outflows, and potential dynamical ejections in massive star-forming regions. Confirming the role of source \#11, whether as the primary ionizing source or as a product of the explosive outflow, will require additional multiwavelength observations and more comprehensive statistical analyses of the molecular outflows. These steps will shed further light on the origin of \exofs\ and on the structure and evolution of \dr.

\section*{Acknowledgements}
We thank E.~Guzmán Ccolque, who provided the positions of the CO outflows detected with ALMA. 
We thank A. Karska for her valuable comments and suggestions which helped us to improve the quality of the manuscript.
V.Y. acknowledges financial support from CONAHCyT, Mexico,from the UNAM-PAPIIT BG100223 grants.
S.A.D. acknowledges the M2FINDERS project from the European Research
Council (ERC) under the European Union's Horizon 2020 research and innovation programme (grant No 101018682).
A.P. acknowledges financial support from the Sistema Nacional de Investigadores of
CONAHCyT, and from the CONAHCyT project number 86372 of the
`Ciencia de Frontera 2019’ program, entitled `Citlalc\'oatl: A
multiscale study at the new frontier of the formation and early
evolution of stars and planetary systems’, M\'exico.
L.A.Z. acknowledges financial support from UNAM-PAPIIT IN112323 grant, México.
We gratefully acknowledge financial support provided by
  Dirección General de Asuntos del Personal Académico,
  Universidad Nacional Autónoma de México,
through grants IG100223 (V.Y. and A.P.) and IN109823 (W.J.H.) of the 
``  Programa de Apoyo a Proyectos de Investigación
  e Inovación Tecnológica''.

The National Radio Astronomy Observatory is a facility of the National Science Foundation operated
under cooperative agreement by Associated Universities, Inc.
This document was prepared using the collaborative tool Overleaf available at 
\url{https://www.overleaf.com/}. 

In addition to the software mentioned in the paper, we have also used the following software: 
Astropy\footnote{\url{https://www.astropy.org/}} \citep{astropy2018}, NumPy\footnote{\url{https://www.numpy.org/}} \citep{numpy2011}, SciPy\footnote{\url{https://www.scipy.org/}} \citep{scipy2014}, Matplotlib\footnote{\url{https://matplotlib.org/}} \citep{matplotlib2007}, and APLpy \citep{aplpy2012}.

%%%%%%%%%%%%%%%%%%%%%%%%%%%%%%%%%%%%%%%%%%%%%%%%%%
\section*{Data Availability}

The data underlying this article will be shared on reasonable request to the corresponding author.



\bibliographystyle{mnras}
\bibliography{references} % if your bibtex file is called example.bib

%%%%%%%%%%%%%%%%%%%%%%%%%%%%%%%%%%%%%%%%%%%%%%%%%%

%%%%%%%%%%%%%%%%% APPENDICES %%%%%%%%%%%%%%%%%%%%%

\appendix

\section{Additional properties of compact radio sources}
\label{app:additional-properties}
We have reported the detection of 13 CRS. Some of the source properties measured in the radio images are
given in the main text, which were important for the discussion. 
In this section, we provide additional measured properties obtained from the different images to complement the work done.

\begin{landscape}%
\begin{table}
%\begin{sidewaystable*}
\begin{minipage}{\linewidth}
\setlength{\tabcolsep}{2.5pt}
\centering
\scriptsize
\footnotesize
%\small
 \caption{Compact radio sources in \dr.}

    \begin{tabular}{ccccccccccccccccccccccccc}
         \hline
         \hline
\# & R.A. & Dec. & 
 {$S_{\rm int, LSB}$}& $S_{\rm peak,LSB}$ &  {$S_{\rm int, USB}$}& $S_{\rm peak,USB}$ & 
{$S_{\rm int,1, LSB}$}& $S_{\rm peak,1,LSB}$ &  {$S_{\rm int,1, USB}$}& $S_{\rm peak,1,USB}$ & $\alpha_1$ & {$S_{\rm int,2, LSB}$}& $S_{\rm peak,2,LSB}$ &  {$S_{\rm int,2, USB}$}& $S_{\rm peak,2,USB}$ &$\alpha_2$ \\
 &$20^{\rm h}39^{\rm m}[^{\rm s}]$&$42^{\circ}19'['']$&
($\mu$Jy) & ($\mu$Jy bm$^{-1}$)&($\mu$Jy) & ($\mu$Jy bm$^{-1}$)&($\mu$Jy) & ($\mu$Jy bm$^{-1}$)& 
($\mu$Jy) & ($\mu$Jy bm$^{-1}$)& &($\mu$Jy) & ($\mu$Jy bm$^{-1}$)&($\mu$Jy) & ($\mu$Jy bm$^{-1}$)& \\
 \hline
1& 0.042& 36.48& $230\pm 18$& $248\pm 11$& $253\pm 18$& $228\pm 9$& $201\pm 16$& $207\pm 9$& $205\pm 24$& $195\pm 13$& $-0.3\pm 0.4$& $274\pm 37$& $311\pm 23$& $301\pm 32$& $263\pm 16$& $-0.8\pm 0.5$\\
2& 0.369& 30.62& $98\pm 33$& $98\pm 18$& $124\pm 28$& $119\pm 15$& $151\pm 35$& $153\pm 20$& $167\pm 41$& $185\pm 23$& $0.9\pm 0.9$& ...& ...& ...& ...& ...\\
3& 0.437& 58.69& $120\pm 23$& $109\pm 12$& $128\pm 30$& $105\pm 15$& $142\pm 12$& $138\pm 6$& $124\pm 43$& $105\pm 22$& $-1.4\pm 1.1$& $90\pm 28$& $95\pm 16$& $148\pm 59$& $120\pm 29$& $1.2\pm 1.5$\\
%4& 0.554& 46.16& $110\pm 14$& $114\pm 8$& $106\pm 31$& $68\pm 13$& $54\pm 17$& $66\pm 11$& ...& ...& ...& $202\pm 25$& $161\pm 12$& $146\pm 30$& $156\pm 18$& $-0.2\pm 0.7$\\
%5& 0.848& 37.04& $227\pm 52$& $153\pm 23$& $210\pm 57$& $139\pm 24$& $211\pm 37$& $177\pm 18$& $226\pm 65$& $155\pm 28$& $-0.7\pm 1.0$& $276\pm 88$& $181\pm 37$& $187\pm 59$& $143\pm 27$& $-1.2\pm 1.4$\\
4& 0.869& 38.89& $261\pm 50$& $197\pm 23$& $248\pm 62$& $181\pm 28$& $423\pm 43$& $230\pm 16$& $219\pm 74$& $184\pm 36$& $-1.1\pm 1.0$& $229\pm 78$& $215\pm 43$& $334\pm 76$& $189\pm 29$& $-0.6\pm 1.3$\\
%7& 0.904& 31.60& $84\pm 29$& $89\pm 17$& $118\pm 49$& $93\pm 23$& $119\pm 37$& $106\pm 19$& $90\pm 53$& $70\pm 25$& $-2.1\pm 2.0$& $207\pm 41$& $155\pm 19$& $267\pm 95$& $110\pm 28$& $-1.7\pm 1.4$\\
5& 1.102& 35.26& $395\pm 43$& $242\pm 17$& $335\pm 58$& $223\pm 24$& $478\pm 66$& $251\pm 24$& $313\pm 67$& $219\pm 29$& $-0.7\pm 0.8$& $403\pm 43$& $277\pm 19$& $351\pm 118$& $157\pm 37$& $-2.8\pm 1.2$\\
6& 1.165& 38.50& $8730\pm 540$& $2690\pm 130$& $8440\pm 540$& $2000\pm 100$& $10310\pm 670$& $2980\pm 150$& $9850\pm 660$& $2380\pm 130$& $-1.1\pm 0.4$& $9150\pm 570$& $2970\pm 140$& $9650\pm 560$& $2160\pm 100$& $-1.6\pm 0.3$\\
7& 1.190& 18.64& $108\pm 15$& $108\pm 8$& $95\pm 22$& $82\pm 11$& $141\pm 13$& $140\pm 7$& $113\pm 37$& $121\pm 21$& $-0.7\pm 0.9$& ...& ...& ...& ...& ...\\
8& 1.215& 35.51& $396\pm 28$& $382\pm 16$& $364\pm 43$& $330\pm 23$& $441\pm 14$& $399\pm 7$& $344\pm 60$& $335\pm 32$& $-0.9\pm 0.5$& $412\pm 60$& $421\pm 34$& $352\pm 65$& $324\pm 35$& $-1.3\pm 0.7$\\
9& 1.234& 37.37& $342\pm 44$& $228\pm 19$& $393\pm 63$& $163\pm 19$& $383\pm 72$& $235\pm 30$& $320\pm 76$& $144\pm 25$& $-2.4\pm 1.1$& $390\pm 57$& $258\pm 25$& $308\pm 67$& $174\pm 26$& $-2.0\pm 0.9$\\
10& 1.329& 34.33& $594\pm 47$& $568\pm 26$& $654\pm 57$& $509\pm 27$& $613\pm 43$& $595\pm 24$& $669\pm 60$& $529\pm 28$& $-0.6\pm 0.3$& $581\pm 59$& $589\pm 34$& $531\pm 71$& $498\pm 39$& $-0.8\pm 0.5$\\
%14& 1.403& 40.31& $396\pm45 155$& $122\pm 37$& $164\pm 59$& $117\pm 27$& $488\pm 120$& $180\pm 34$& $385\pm 214$& $112\pm 49$& $-2.4\pm 2.4$& $230\pm 89$& $130\pm 34$& $1340\pm 360$& $98\pm 25$& $-1.4\pm 1.8$\\
11& 1.475& 34.61& $1590\pm 110$& $1025\pm 45$& $1670\pm 100$& $903\pm 39$& $1481\pm 91$& $1004\pm 40$& $2120\pm 150$& $952\pm 48$& $-0.3\pm 0.3$& $2190\pm 130$& $1197\pm 50$& $1400\pm 120$& $919\pm 50$& $-1.3\pm 0.3$\\
%16& 1.712& 38.07& $249\pm 64$& $124\pm 22$& $121\pm 53$& $68\pm 19$& $325\pm 51$& $172\pm 18$& $240\pm 101$& $77\pm 24$& $-4.0\pm 1.6$& $385\pm 161$& $117\pm 38$& $259\pm 115$& $85\pm 28$& $-1.6\pm 2.3$\\
12& 1.733& 36.40& $596\pm 44$& $417\pm 20$& $625\pm 70$& $338\pm 26$& $672\pm 45$& $450\pm 20$& $709\pm 67$& $361\pm 24$& $-1.1\pm 0.4$& $652\pm 68$& $439\pm 30$& $574\pm 78$& $366\pm 33$& $-0.9\pm 0.6$\\
13& 1.825& 39.01& $408\pm 52$& $266\pm 22$& $465\pm 66$& $194\pm 20$& $521\pm 80$& $253\pm 27$& $562\pm 124$& $200\pm 33$& $-1.2\pm 1.0$& $355\pm 58$& $282\pm 28$& $355\pm 58$& $198\pm 22$& $-1.8\pm 0.7$\\
 
         \hline
    \end{tabular} \label{tab:crsA}\\ 
%\end{sidewaystable*}
\end{minipage}
\end{table}
  
\end{landscape}

\section{Algorithm for fitting a parabola or other conic section to a set of points}
\label{app:pap}

We here present an algorithm for fitting a conic section curve
(ellipse, parabola, hyperbola) to a set of points \((x_i, y_i)\),
\(i \in \{1, \dots, N\}\),
which lie in a two-dimensional plane.
This is implemented in the \textsc{confitti} python library,
which is freely available from the Python Package Index\footnote{%
  \url{https://pypi.org/project/confitti} or \texttt{pip install confitti}}
or from GitHub.\footnote{%
  \url{https://github.com/div-B-equals-0/confitti}}
This description corresponds to version 0.2.1 of the library.
The algorithm is a spiritual successor to the \textsc{circle-fit} method
described in Appendix~E of \citet{Tarango-Yong:2018a}.

\subsection{Conic section parameters}
\label{sec:conic-sect-param}

\begin{figure}[!ht]
    \centering
    \includegraphics[width=0.9\linewidth]{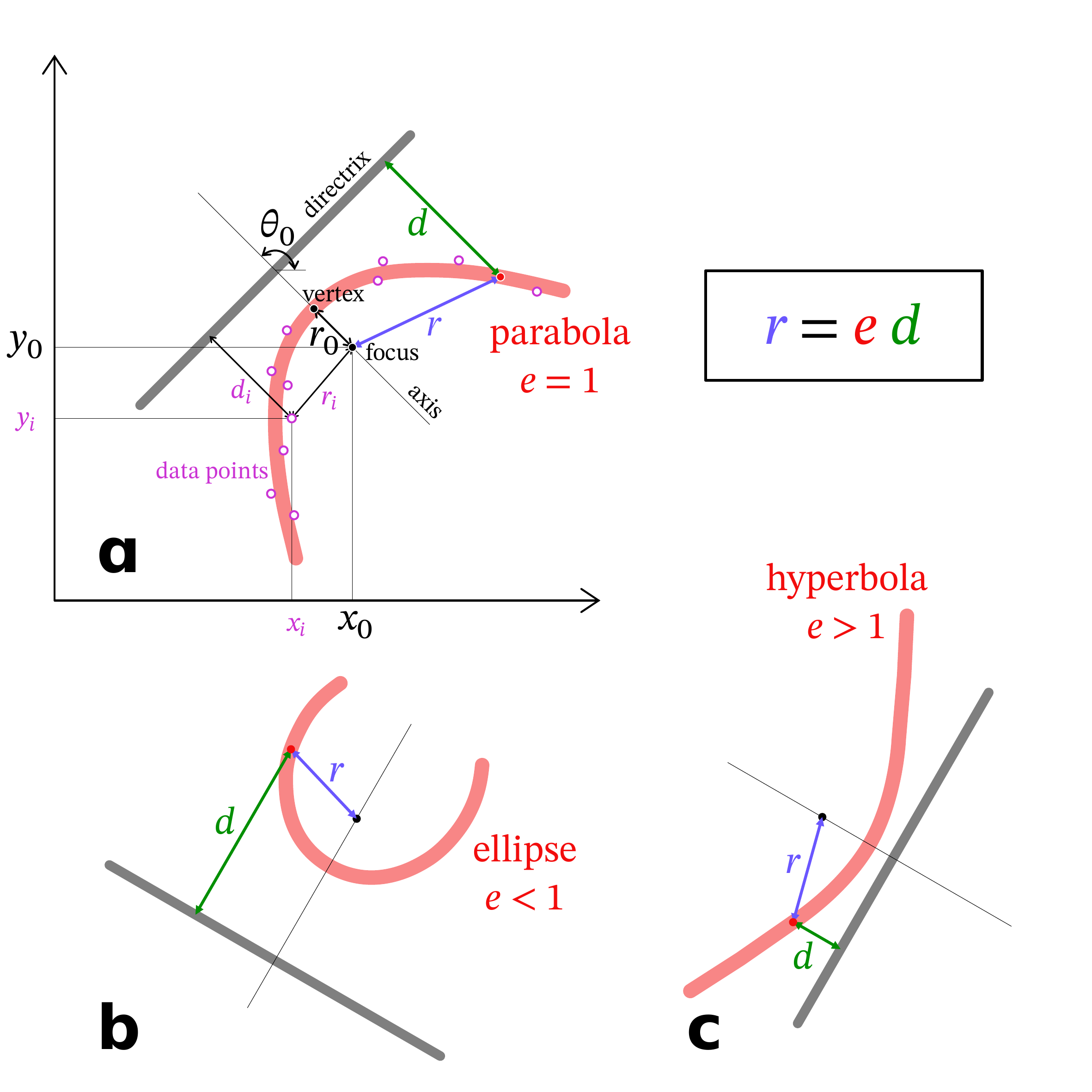}
    \caption{
      Illustration of the procedure for fitting conic sections
      (thick red curves) to a set of observed points (small open circles)
      by making use of the focus-directrix relation: \(r = e d\).
      (a)~Case of parabola (\(e = 1\)),
      labeled with all the relevant quantities as described in the text.
      (b)~Case of ellipse (\(e < 1\)).
      Note that only a portion of the ellipse is shown.
      (c)~Case of hyperbola (\(e > 1\)).
    }
    \label{fig:parS}
\end{figure}

A conic section curve can be completely specified by
a set of 5 parameters: \(\{x_0, y_0, r_0, \theta_0, e\}\), where
\((x_0, y_0)\) is the position of the focus,
\(r_0\) is the size,
\(\theta_0\) is the orientation angle of the curve's symmetry axis,
and \(e\) is the eccentricity of the conic section.
Hyperbolae have \(e > 1\), a parabola has \(e = 1\),
while ellipses have \(e < 1\),
with \(e = 0\) corresponding to a circle
(in which case \(\theta_0\) is undefined).
The size \(r_0\) is taken to be the distance between the focus and the apex
(see Figure~\ref{fig:parS}),
which is the point on the curve that is closest to the focus.
The symmetry axis passes through both the focus and the apex,
with \(\theta_0\) being measured anticlockwise from the \(x\) axis.
% reference direction (for instance, celestial North).

\subsection{Minimization of focus-directrix residuals}
\label{sec:minim-focus-directr}

The algorithm makes use of the focus-directrix property of conic section curves
\citetext{e.g., \citealp{Brannan:2012a} \S~1.1.3}:
\begin{equation}
  \label{eq:focus-directrix}
  r = e d,
\end{equation}
where \(r\) is the distance from the focus to a point \(P\) on the curve,
and \(d\) is the perpendicular distance of the point \(P\) from the directrix.
The directrix is a line perpendicular to symmetry axis that crosses the axis
at a distance \((1 + e^{-1})\, r_0\) from the focus (see Figure~\ref{fig:parS}).
For a general point \((x_i, y_i)\),
which does not necessarily lie on the curve,
we can calculate a residual difference \(\delta_i = r_i - e d_i\), where
\begin{equation}
  \label{eq:data-point-r}
  r_i = \left( (x_i - x_0)^2 + (y_i - y_0)^2\right)^{1/2}
\end{equation}
and
\begin{equation}
  \label{eq:data-point-d}
  d_i = (1 + e^{-1})\, r_0 -  (x_i - x_0) \cos\theta_0 - (y_i - y_0) \sin\theta_0 .
\end{equation}
We can then define the best-fitting conic section curve as the
set of parameters \(\{x_0, y_0, r_0, \theta_0, e\}\) that minimizes
the sum of squared, scaled residuals:
\begin{equation}
  \label{eq:sum-squares}
  S = \sum_{i = 1}^N (\delta_i / \sigma_i)^2 ,
\end{equation}
where the scaling is by the uncertainties \(\sigma_i\) in the \((x_i, y_i)\) measurements (if available).

\subsection{Implementation details}
\label{sec:impl-deta}

The \textsc{confitti} library provides the function 
\texttt{fit\_conic\_to\_xy\allowbreak (...)}
with arguments \texttt{xpts} and \texttt{ypts}, which correspond to the
observed data points \((x_i, y_i)\) discussed above.
This function takes care of making an initial guess for the
conic section curve parameters, and then refining this guess by
minimizing equation~(\ref{eq:sum-squares})
using the Levenberg-Marquardt algorithm \citep{More:1978a} as implemented by the
\textsc{scipy} \citep{Virtanen:2020a}
and \textsc{lmfit} \citep{matt_newville_2024_12785036} Python libraries.
% \footnote{\url{https://lmfit.github.io/lmfit-py}}
The initial guess for the parameters is as follows:
\begin{enumerate}
\item The focus point \((x_0, y_0)\) is initialized to the mean position of the data points.
\item The size \(r_0\) is initialized to the mean distance of the 5 closest data points from the initial focus.
\item The orientation \(\theta_0\) is initialized to the circular mean angle
  of the 5 closest data points from the initial focus.
\item The eccentricity is initialized to \(e = 1\).
\end{enumerate}
Extensive testing shows that this method works well,
so long as the observed data points cover both sides of the apex.
Typically, 20 to 50 evaluations of the residual function are required
to converge the solution to the default relative tolerance of \(\sim 10^{-8}\). 

Optionally, the fit may be constrained in various ways.
For example, the \texttt{only\_parabola} option
freezes the eccentricity at \(e = 1\), such that the fitted
curves have only 4 free parameters instead of 5.
Explicit per-point uncertainties in the data points,
\(\sigma_i\), may also be provided, but by default these are set to unity.
Additionally the library provides functions for saving and loading the fit results in JSON and YAML formats,
and for plotting the fitted curves.
All features are demonstrated in example jupyter notebooks.\footnote{%
  \url{https://github.com/div-B-equals-0/confitti/tree/main/notebooks}
}

\subsection{Quantifying uncertainty in derived parameters}
\label{sec:quant-uncert-deriv}
In addition to the estimated best-fit parameters,
the Levenberg-Marquardt minimization provides a \(\chi^2\) measure
of the goodness of fit\footnote{%
  The absolute value of \(\chi^2\) is only meaningful in the case that
  explicit uncertainties \(\sigma_i\) in the data points are provided.
}
together with a covariance matrix that expresses the estimated
uncertainties in the fitted parameters and the correlations between these.
Alternatively, the posterior probability distribution of the fitted curve parameters
can be explored more thoroughly through 
Markov Chain Monte Carlo (MCMC) ensemble sampling \citep{Goodman:2010a}
as implemented via \textsc{lmfit}'s interface to  the
\textsc{emcee} Python library \citep{Foreman-Mackey:2013a}.
By default, we assume uniform priors for all the parameters.
In the case that the data-point observational errors \(\sigma_i\)
are not known a priori,
then a uniform error \(\ln \sigma\) can be added as an additional
parameter to the model,
allowing the posterior probability distribution of the data errors
to be estimated also.
Note that the MCMC calculation is relatively computationally expensive,
requiring of order \(5 \times 10^5\) function evaluations,
which is an increase in effort of 4 orders of magnitude,
compared with the Levenberg-Marquardt minimization.

\begin{figure}
  \centering
  \includegraphics[width=\linewidth]{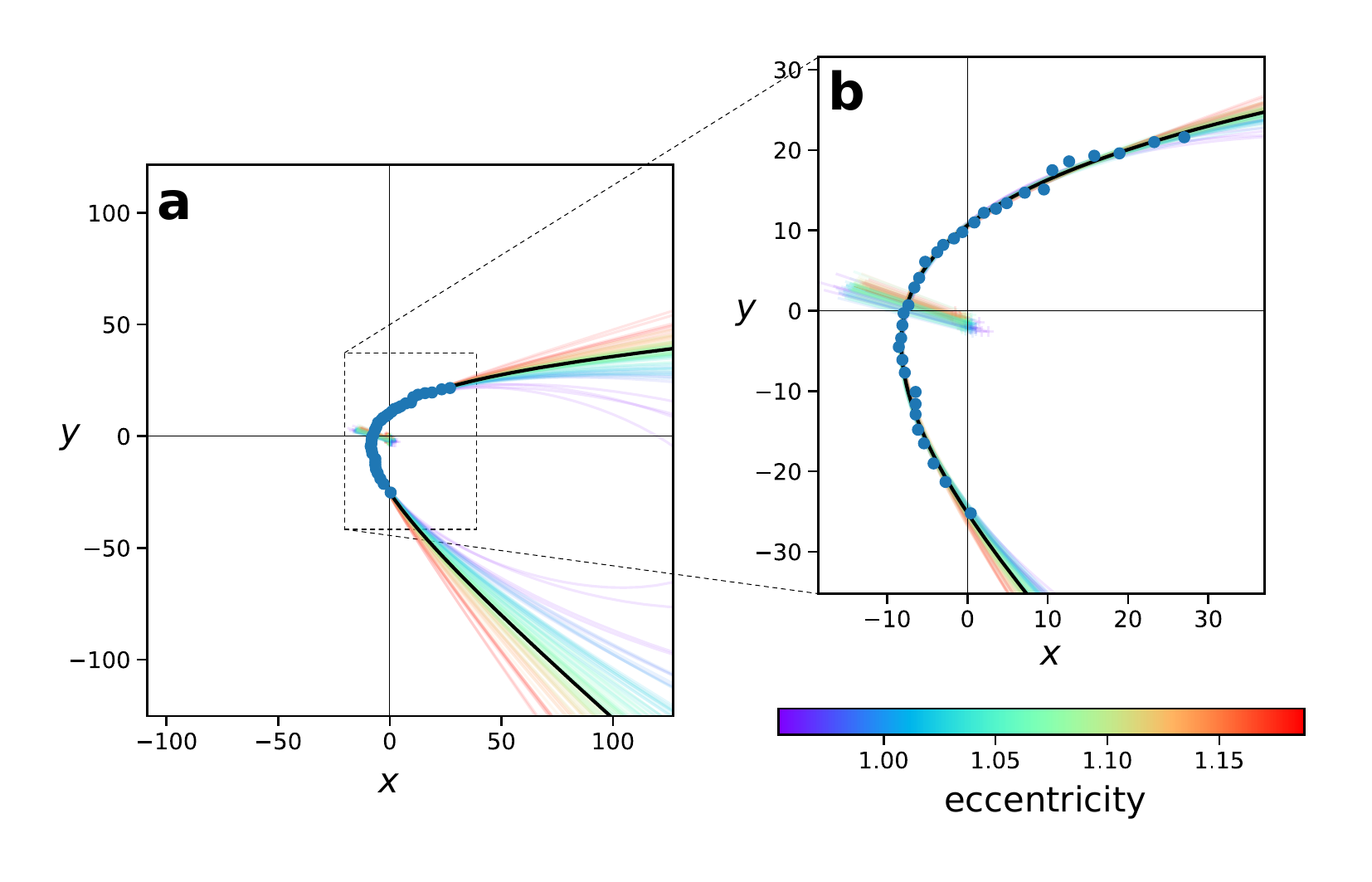}
  \caption{Example fit of conic section arc to data points (blue symbols).
    (a) Zoomed out view. (b) Zoomed in view.
    Black line shows best-fit curve.
    Colored lines show a random sample of curves
    from the MCMC posterior distribution of the conic parameters,
    colored according to the eccentricity (ellipses in purple,
    parabola in blue, hyperbolae in green-yellow-red).
    Also shown are the focus point of each curve (plus symbols)
    and a straight line that indicates the portion of the symmetry axis
    between the focus and the directrix,
    both colored in the same way.
  }
  \label{fig:example-fit}
\end{figure}

\begin{figure}
  \centering
  \includegraphics[width=\linewidth]{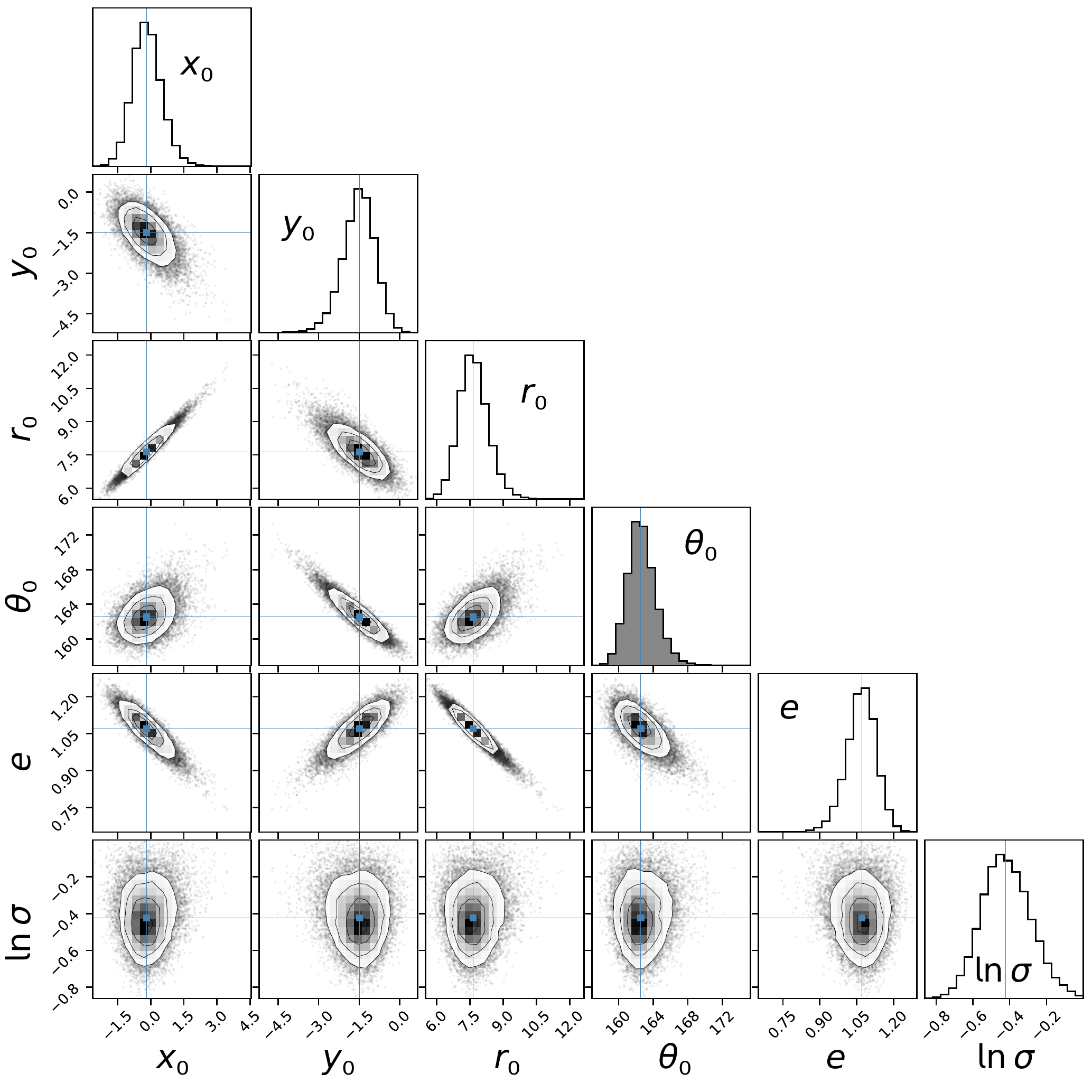}
  \caption{
    Corner plot of MCMC-derived covariances between fitted model parameters
    from the same example as shown in Figure~\ref{fig:example-fit}.
    Plots on the leading diagonal show the one-dimensional histogram of
    the posterior distribution of each parameter,
    assuming a uniform prior distribution.
    Off-diagonal plots show the two-dimensional histogram of
    the joint posterior distribution of each pair of parameters.
    Blue dots and lines indicate the best-fit parameters.
    The bottom row shows the natural logarithm of the estimated
    magnitude of observational errors for the individual data points
    (assumed constant for all points), which is included as an
    additional parameter in the model.   
  }
  \label{fig:corner-plot}
\end{figure}

An example is shown in Figure~\ref{fig:example-fit}, where colored lines
show curves with parameters taken from a random selection of
100 posterior samples from the MCMC chain,
which gives an estimate of the uncertainty around the best-fit curve
(black line).
Figure~\ref{fig:corner-plot} shows the pairwise correlations
and marginal histograms of the posterior distributions of conic parameters
from the same example fit,
expressed as a corner plot \citep{Foreman-Mackey:2016a}.

\subsection{Comparison with previous work}
\label{sec:comp-with-prev}

  In this section we justify the focus-directrix approach adopted
  by the \textsc{confitti} library by comparing its performance with other
  algorithms from the literature.

A variety of different approaches to conic fitting are described in \citet{Zhang:1997b}, with the two main categories being
algebraic and geometric methods.
In the algebraic approach, the conic curve is a solution of the
quadratic equation:
\begin{equation}
  \label{eq:algebraic}
  A x^2 + 2 B x y + C y^2 + 2 D x + 2 E y + F = 0,
\end{equation}
together with a normalization constraint.
For instance, with the normalization \(F = 1\), then
the curve is specified by the parameters \(\{A, B, C, D, E\}\).
One disadvantage of algebraic methods is
that data points in high-curvature parts
of the conic have less influence on the solution than
points in low-curvature parts.

  This is especially important for the application to the DR~21 arcs (section~\ref{sec:arc-shap-struct-results})
  since the high-curvature parts are precisely those near
  the apex of the parabola,
  which are crucial in constraining the orientation
  of the axis (section~\ref{sec:arc-shap-struct} and Fig.~\ref{fig:all_lines}).

A second disadvantage is that the parameters \(\{A, B, C, D, E\}\)
transform in a complicated way
under translations and rotations of the axes,
unlike our preferred parameters
\(\{x_0, y_0, r_0, \theta_0, e\}\) (section~\ref{sec:conic-sect-param}),
which have a more intuitive interpretation in terms of the conic shape.

The most obvious geometric method is to minimize the sum of
squared orthogonal distances of the points from the curve
\citetext{\citealp{Zhang:1997b}, \S~5.2}.
However, computation of the orthogonal distance is complicated
and requires solving a quartic equation.
We therefore prefer to instead make use of the
much simpler focus-directrix relation described above (section~\ref{sec:minim-focus-directr}).
Our method is closely related to the geometric fitting technique
proposed by \citet{Lopez-Rubio:2018a} (LR18),
but with the following key differences.
\begin{enumerate}
\item Our method applies to general conics, while LR18 is restricted to parabolas.
\item LR18 employed a bespoke minimization algorithm,
  with explicit analytic calculation of the gradients of the errors,
  and including multiple
  restarts of the process from a random perturbation of an initial solution.
  In our case we prefer to employ
  a general-purpose off-the-shelf minimization library
  with numerical estimates of the gradients,
  and replacing the random restarts with the option to perform
  MCMC ensemble sampling.
\item LR18 added an additional penalty term to the cost function
  in order to avoid degenerate solutions,
  whereas we find that the same goal can be achieved
  by constraining the conic size to lie within a factor of 3 of the initial guess.
\item LR18 use minimization of absolute errors to improve the robustness of the method,
  whereas we find that regular minimization of mean-square errors is sufficient.
\end{enumerate}

\section{Parabola fit to G5.89} \label{app:gf}

To study the extended emission of \gf, we applied a similar analysis as that used for \dr. As shown in Fig.~\ref{fig:streams}, the shell-like structure in \gf\ does not display well-defined arcs within it, unlike in \dr. However, we identified a possible arc that was used for the fitting procedure. For this analysis, we used the FB image including all baselines and combining the two epochs, as this provides the highest sensitivity to extended emission.

\begin{figure}
    \centering
    \includegraphics[height=0.3\textwidth]{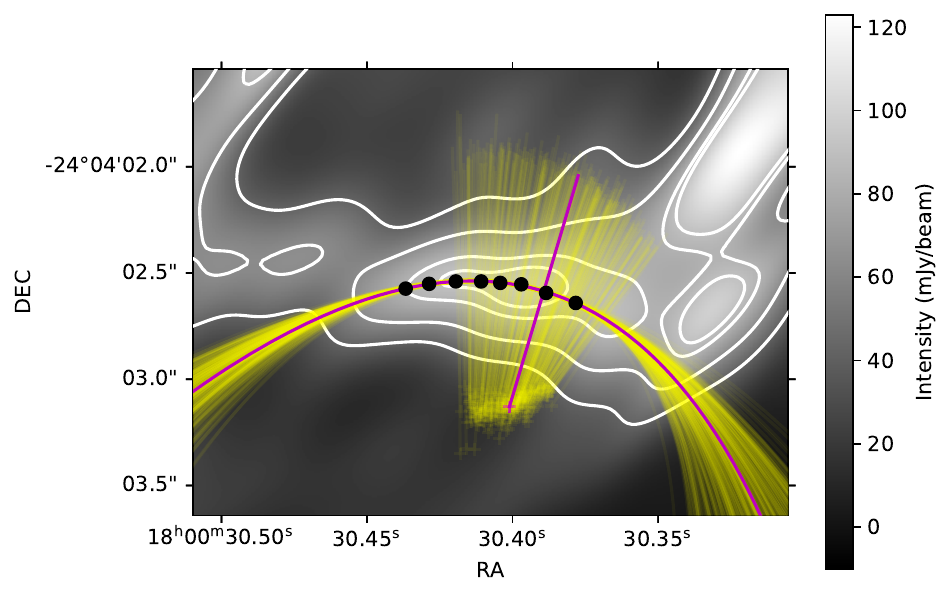}
    \caption{Detailed view of the MCMC parabola fit for a southern arc in the shell of the \gf\ cm image. The black points are the data taken from our combined FB image with all baselines. The magenta parabola shows the best fit and reveals the orientation of the parabola, and the magenta cross presents the parabola focus $x_f$, $y_f$. The yellow lines indicate the different possibilities found using MCMC ensemble sampling. Contours are 60, 90, 120 and 130 times the noise level of the image, 692\,$\mu$Jy~beam$^{-1}$.}
    \label{fig:appgf}
\end{figure}

The best-fit parabola has its focus at $x_f = 18^{\rm h}00^{\rm m}30.40^{\rm s} \pm 0.02^{\rm s}$ and $y_f = -24^{\circ}04'03.09'' \pm 0.01''$, with parameters $r_0 = 0.5'' \pm 0.2''$ and $\theta = -23^\circ \pm 27^\circ$. The result is shown in Figure~\ref{fig:appgf}.

%%%%%%%%%%%%%%%%%%%%%%%%%%%%%%%%%%%%%%%%%%%%%%%%%%

% Don't change these lines
\bsp	% typesetting comment
\label{lastpage}
\end{document}